\documentclass[fleqn,usenatbib]{mnras}
\usepackage{newtxtext,newtxmath}
\usepackage[T1]{fontenc}
\usepackage{ae,aecompl}

\usepackage{graphicx}
\usepackage{amsmath}
\usepackage{amssymb}

\newcommand{\Nside}{\ensuremath{N_{\text{Side}}}}

\title[Statistics of polarized CMB foreground maps]{Statistical properties of polarized CMB foreground maps}

\author[S. von Hausegger et al.]{
Sebastian von Hausegger,$^{1}$\thanks{E-mail: s.vonhausegger@nbi.dk}
Aske Gammelgaard Ravnebjerg$^{1}$
and Hao Liu$^{1,2}$
\\
$^{1}$Niels Bohr Institute \& Discovery Center, University of Copenhagen, Blegdamsvej 17, DK-2100 Copenhagen \O, Denmark\\
$^{2}$Key laboratory of Particle and Astrophysics, Institute of High Energy Physics, CAS, 19B YuQuan Road, Beijing, China
}

\date{Accepted XXX. Received YYY; in original form ZZZ}

\pubyear{2018}

\begin{document}
\label{firstpage}
\pagerange{\pageref{firstpage}--\pageref{lastpage}}
\maketitle

\begin{abstract}
Foreground removal techniques for CMB analyses make specific assumptions about the properties of foregrounds in temperature and in polarization.  By investigating the statistics of foreground components more understanding about the degree to which these assumptions are valid can be obtained.  In this work we investigate $E$- and $B$-mode maps of the two strongest polarized foregrounds, synchrotron and thermal dust emission, with regards to their similarity with Gaussian processes, their spectral variations and cross-correlations.  We perform tests in patches of $\sim3.7^\circ$ size collectively covering the full sky and find most of them to conform with their Gaussian expectation according to the statistics in use.  Correlations exhibit distinct differences in $E$- and $B$-mode signals which point towards necessities in foreground removal methods.  We discuss potential consequences and possible further directions.
\end{abstract}

\begin{keywords}
cosmic background radiation -- cosmology: observations
\end{keywords}

%%%%%%%%%%%%%%%%%%%%%%%%%%%%%%%%%%%%%%%%%%%%%%%%%%

%%%%%%%%%%%%%%%%% BODY OF PAPER %%%%%%%%%%%%%%%%%%

\section{Introduction}
\label{sec:1}

The polarization of the Cosmic Microwave Background (CMB) offers unique insights into the dynamics of light and matter around redshift $z\sim1100$, the epoch of re-ionization around $z\sim6$, and possibly primordial gravitational waves.  (For a recent review of the corresponding physics and experimental achievements, see~\cite{Staggs:2018gvf}.)  This latter contribution is conveniently characterized by the tensor-to-scalar ratio $r$ whose detection is yet outstanding.  While the most recent analyses result in upper bounds of $r<0.07$~\citep{Aghanim:2018eyx,2015PhRvL.114j1301B}, future surveys forecast sensitivities of down to $r\sim10^{-3}$ or even $10^{-4}$ (e.g.~\cite{Abazajian:2016yjj}), based on our current knowledge of those sources interfering with a clean measurement.  Galactic foregrounds and systematics, especially in polarization, seem to offer the greatest challenges, and the corresponding methods for their removal~\citep{Aghanim:2016yuo,Akrami:2018mcd} are perpetually tailored to our continuously improving understanding of their influences.

The Galactic foreground components contributing most to the polarized microwave and millimeter sky are synchrotron radiation, at low, and thermal dust emission, at high frequencies; others are presumably only polarized at the percent level.  Many properties of the polarized synchrotron and thermal dust skies are already under thorough investigation~\citep{Ade:2014zja,Ade:2014gna,Akrami:2018wkt,Krachmalnicoff:2018imw}.  For instance, out of immediate relevance to the extraction of cosmological parameters from the CMB, those properties related to their power spectra have been of great interest~\citep{Adam:2014bub}, of which the surprisingly constant ratio of approximately~2 of the dust's $E$-to-$B$-mode power~\citep{Ade:2015mbc} at intermediate to high multipoles has sparked considerable attention~\citep{Caldwell:2016xkd,Kritsuk:2017aab,Kandel:2017xjx}.  While also of astrophysical interest, statistical investigations of Galactic foregrounds can also assist their very removal to obtain a sufficiently cleaned CMB map.

Searches for and detection of residual foreground emission in CMB products have been practiced already since the first release of the \textit{WMAP} data (e.g.,~\cite{Naselsky:2003tk,Naselsky:2003na,Dineen:2003qc}) and more recently on \textit{Planck} data (e.g.~\cite{vonHausegger:2015flg}), highlighting potential downsides of assumptions made in foreground removal algorithms.  Besides the general hope for reducing residual contamination in the final product, meticulous searches for statistical peculiarities in the CMB~\citep{Ade:2015hxq}, e.g., primordial non-Gaussianity~\citep{Bartolo:2010qu}, must be safe from distortion of these very distributions from contaminants such as systematic noise and/or foregrounds.  Only recently more focus has been placed on understanding the relationship of foreground's statistics with those which are used to search for non-Gaussianity in the final CMB products, see e.g.~\cite{Hill:2018ypf, Jung:2018rgf, Coulton:2019bnz}.  In this work we would like to take a step back and investigate the statistical behavior of foregrounds in a general manner, independent of studies of specific CMB non-Gaussianities, in order to, among others, finally make connections to general issues in foreground removal methods.    Also this has been of recent interest~\citep{Ben-David:2015fia,Rana:2018oft} regarding temperature maps of Galactic synchrotron radiation, which calls for similar studies of maps of polarization.

At present, foreground separation techniques place requirements on certain properties of the foreground components -- some explicitly, such as parametric foreground fitting algorithms~\citep{2008ApJ...676...10E}, and some implicitly, such as so-called blind foreground removal techniques~\citep{Bennett:2003bz,2008ISTSP...2..735C,2009A&A...493..835D}.  By the example of the latter, we shall, in section~\ref{sec:2} of this paper, provide motivation to analyse statistics of these foreground components regarding their Gaussianity and their relations to each other.  The evaluation of all such algorithms is most effectively assisted by a set of simulated foregrounds, which in turn depend on assumptions about their statistics:  The simulation of certain foregrounds' properties is often underpinned by assuming Gaussian behavior~\citep{Tegmark:1999ke,1996MNRAS.281.1297T}, and small-scale structure is commonly added into existing maps by simply generating Gaussian fluctuations according to extrapolations of the available power spectra~\citep{Remazeilles:2014mba,Hervias-Caimapo:2016crc}.  As accurate simulations are instrumental in making realistic predictions for forthcoming surveys, exploiting possible Gaussian patterns in foregrounds' statistics will prove itself useful.  In other words, understanding foregrounds' statistics will enable to justify (and expand on) assumptions in current foreground simulation procedures, such as those of~\cite{Thorne:2016ifb} or~\cite{Delabrouille:2012ye}.  Some recent effort has already gone into the determination at which scales the assumption of Gaussianity in foregrounds holds~\citep{Ben-David:2015fia,Rana:2018oft}.  These findings are to be extended to studies of polarization which for the first time is attempted in this work, by the investigation of Galactic foreground $E$- and $B$-mode maps.

In short, the aim of this paper is to elucidate assumptions inherent to current foreground removal techniques and to, once convinced that foreground residuals will be present in CMB maps at some level, generally characterize their statistics.\\

After a general motivation in section~\ref{sec:2}, our procedure for classifying Gaussianity is briefly reviewed in section~\ref{sec:3}.
In section~\ref{sec:4} we then present the results from $E$- and $B$-mode maps at 23 and 353~GHz, corresponding to the polarization signal from synchrotron and from thermal dust emission, respectively.  We conclude in section~\ref{sec:5}.

\section{Motivation}
\label{sec:2}

To illustrate the necessity of studying the statistics of foregrounds, we provide the following motivation.  A commonly used framework to understand so-called blind foreground subtraction algorithms is that of Internal Linear Combination, or ILC, which has been introduced to CMB science by the \textit{WMAP} collaboration~\citep{Bennett:2003bz}.  Sky maps of different frequency are linearly combined with the aim to arrive at a map of the CMB anisotropy.  More sophisticated algorithms are in use today, see e.g.~\cite{Adam:2015tpy}, albeit they produce CMB products consistent with an ILC solution.  In this section we shall briefly review essential elements and assumptions of the ILC method, as a representative of such blind foreground removal techniques.

The method goes as follows~\citep{Eriksen:2004jg}.  Consider a signal, $s_\nu(p)$, on the sky consisting of only CMB anisotropy, $c(p)$, and a term describing the foreground, $f_\nu(p)$, for each pixel $p$, measured at frequency $\nu$, and in units where the CMB's contribution is frequency independent.  A weighted sum over all frequency maps reads
\begin{align}
\mathcal{S}(p)=\sum_\nu w_\nu S_\nu(p) = C(p) + \sum_\nu w_\nu F_\nu(p),
\label{eq:1}
\end{align}
for which we assumed $\sum_\nu w_\nu = 1$ and where we defined
\begin{align*}
&S_\nu(p) = s_\nu(p)-\langle s_\nu\rangle_\Omega,\\
&C(p) = c(p) - \langle c\rangle_\Omega,\\
&F_\nu(p) = f_\nu(p) - \langle f_\nu\rangle_\Omega,
\end{align*}
and the angular brackets $\langle...\rangle_\Omega$ denote the average over all pixels $p\in\Omega$, the region of the sky under consideration.  The incentive of the ILC method is to fix the weights $w_\nu$ such that the last term in Eq.~(\ref{eq:1}) vanishes.  This is not generally possible, and is conventionally solved by requiring the variance of $\mathcal{S}$ to be minimal.  The variance of Eq.~(\ref{eq:1}) then reads
\begin{align}
\langle \mathcal{S}^2 \rangle_\Omega = \langle C^2 \rangle_\Omega + 2\sum_\nu \langle C\, F_\nu \rangle_\Omega \, w_\nu + \sum_{\nu\mu} \langle F_\nu\, F_\mu \rangle_\Omega \, w_\nu w_\mu.
\label{eq:2}
\end{align}
The first term in Eq.~(\ref{eq:2}) corresponds to the variance of the CMB signal, the second describes the chance correlations between the CMB and the total foreground in the frequency band $\nu$, and the last term depends on the cross-correlation matrix of the foregrounds at frequencies $\nu$ and $\mu$.  All these quantities are evaluated in the entire region $\Omega$.\\

To understand the construction of this method we first consider the simplified case in which we can express the foreground term as
\begin{align}
F_\nu(p) = \alpha_\nu F(p),
\label{eq:3}
\end{align}
where $\alpha_\nu=const.$ and scales the template $F(p)$ along frequencies.  In the case of a foreground with power-law emission, $\alpha_\nu = (\nu/\nu_0)^\beta$, the spectral index $\beta=const.$ across the entire region $\Omega$.  From Eq.~(\ref{eq:1}) it can be seen that the CMB is solved for by computing the weights as
\begin{align}
\sum_\nu\alpha_\nu w_\nu = 0,
\label{eq:4}
\end{align}
which at the same time\footnote{It should be clear that in the case where Eq.~(\ref{eq:3}) (or its extension to $n$ foregrounds) is fulfilled, the minimization of the variance becomes redundant.  In the case of a single foreground only two observations at different frequencies are required to solve for the weights $w_\nu$:
\begin{align*}
w_1=-\frac{\alpha_2}{\alpha_1}\left(1-\frac{\alpha_2}{\alpha_1}\right)^{-1},\hspace{0.2cm} w_2= \left(1-\frac{\alpha_2}{\alpha_1}\right)^{-1}
\end{align*}}
 minimizes the variance, Eq.~(\ref{eq:2}).  This even generalizes to the case where $F$ describes the superposition of $n$ foregrounds, each scaled similar to Eq.~(\ref{eq:3}).  Such a system is fully determined for observations at $n+1$ frequencies, or more.\\

In reality, Galactic foregrounds do not follow Eq.~(\ref{eq:3}) perfectly, i.e.~$\alpha_\nu$ must be promoted to be direction dependent, or $\alpha_\nu = \alpha_\nu(p)$.  In terms of a power-law foreground, this translates into a spatially varying spectral index.\footnote{Indeed, with the onset of more precise measurements of the radio sky, it will become increasingly clear that the spectral index of synchrotron emission varies across the sky~\citep{Taylor:2018uaw,Krachmalnicoff:2018imw}, and also for thermal dust emission such variation has been observed and is known as de-correlation~\citep{Aghanim:2016cps} (see, however,~\cite{Sheehy:2017gfx} and~\cite{Akrami:2018wkt})}  The minimization of Eq.~(\ref{eq:2}) then results in $w_\nu$ such that there are pixels, $p$, for which $\sum_\nu w_\nu F_\nu(p)=0$ does not hold.  In other words, residual emission in the final CMB product becomes inevitable with this method.
We expand on this point and consider the factor $\alpha_\nu\rightarrow\alpha_\nu\cdot(1+\Delta_\nu(p))$, where $\Delta_\nu(p)$ can be assumed Gaussian.  We find that the equivalent of Eq.~(\ref{eq:4}),
\begin{align}
\sum_\nu\alpha_\nu w_\nu(1+\Delta_\nu(p)) = 0,
\label{eq:5}
\end{align}
now cannot be satisfied within the entire region $\Omega$ up to the Gaussian term $\sum_\nu\alpha_\nu w_\nu\Delta_\nu(p)$.  For the same reason the variance of $\mathcal{S}$ is biased by the term
\begin{align}
\left\langle F^2(p)\sum_{\nu\mu} \left[2 + \Delta_\nu(p)\right]\Delta_\mu(p)\alpha_\nu\alpha_\mu w_\nu w_\mu \right\rangle_\Omega,
\label{eq:6}
\end{align}
in addition to the one describing chance correlations between the CMB and $\Delta_\nu(p)$.  The first term accounts for the correlation between $\Delta_\nu(p)$ and the foreground `intensity', $F^2(p)$, while the second term, $\propto (F(p)\Delta_\nu(p))^2$, scales like the variance of the $F$--$\Delta$ correlations.  In the case of a power-law foreground with $\alpha_\nu(p)=(\nu/\nu_0)^{\beta(p)}$ the spectral index can be written as $\beta(p)=\beta+\delta\beta(p)$.  Assuming that the variation $|\delta\beta(p)|\ll\beta$ we can approximate $\alpha_\nu(p)\approx\alpha_\nu(1+\delta\beta(p)\ln(\nu/\nu_0))$ which leads to the corresponding form of Eq.~(\ref{eq:6}),
\begin{align}
\left\langle F^2(p)\sum_{\nu\mu} \left[2 + \delta\beta(p)\ln\left(\frac{\nu}{\nu_0}\right)\right]\delta\beta(p)\ln\left(\frac{\mu}{\nu_0}\right)\alpha_\nu\alpha_\mu w_\nu w_\mu \right\rangle_\Omega.
\label{eq:7}
\end{align}
\\

Keeping in mind the potentially varying properties of foregrounds across the sky, we allow ourselves one last remark about the ILC approach, where we consider those patches of the sky, in which the foregrounds are completely uncorrelated from band to band.  In these zones one can model the correlation matrix $\langle F_\nu\, F_\mu\rangle_\Omega$ in Eq.~(\ref{eq:2}) as a diagonal matrix:
\begin{align}
\langle F_\nu\, F_\mu\rangle_\Omega = \langle F_\nu\, F_\mu\rangle_\Omega\,\delta_{\nu,\mu}
\label{eq:8}
\end{align}
where $\delta_{\nu,\mu}$ is the Kronecker symbol. Then, the contribution to the total variance $\langle\mathcal{S}^2\rangle_\Omega$ from the foreground components simplifies to
\begin{align}
    \sum_{\mu\nu}\langle F_\nu\, F_\mu\rangle_\Omega \, w_\nu w_\mu = \sum_\nu \langle F^2_\nu\rangle_\Omega\, w^2_\nu,
    \label{eq:9}
\end{align}
which is overall positive-definite.  Already given Eq.~(\ref{eq:8}), we see that foregrounds of this sort cannot be removed by linear combination.\\

In this rough exploration we have not considered noise terms.  However, their inclusion would only lead to the addition of the noise covariance matrix, and the corresponding cross-covariance matrices in Eq.~(\ref{eq:2}), and not change the arguments about the foreground properties made above.  In particular, Eqs.~(\ref{eq:8})--(\ref{eq:9}) look equivalent for non-correlated noise terms.  Yet there is an essential difference between uncorrelated foreground and uncorrelated noise:  By the continuing improvement in detector technology, one may obtain lower noise levels and thereby reduce the influence of this term, the contribution from uncorrelated foregrounds, however, remains the same.  In the case of \textit{Gaussian}, uncorrelated foregrounds (or noise) Eq.~(\ref{eq:8}) completely describes their properties, and in the light of the previous discussion we find Gaussianity in a region $\Omega$ to arise from either the foreground `template' $F(p)$ itself, or from the direction-dependent term $\Delta_\nu(p)$, while the respective other remains nearly constant.  In particular, any residual of the sort described in Eq.~(\ref{eq:5}), will itself be Gaussian.  The later distinction of such contamination from the also Gaussian CMB will be challenging.

It is these considerations which motivate a spatially resolved investigation of foregrounds with regards to their statistics -- in specific, we shall investigate their similarity (or dissimilarity) to Gaussian variables in small patches distributed over the sky, in addition to studying correlations between foreground maps on the same scales.  In a previous study~\citep{Ben-David:2015fia} we have performed such tests on a full-sky temperature map at 408~MHz, representative for Galactic synchrotron radiation; we here extend the analysis to polarized foregrounds.  Even though formulated for temperature fields, above considerations hold also for the removal of polarization foregrounds.\footnote{In addition to the assumptions about constant spectral indices or the like, the ILC approach in polarization requires the polarization angles of each component to be constant across frequencies.  It is not clear whether the ILC in polarization is best performed on the Stokes parameters $Q$ and $U$, or their non-local transformations $E$ and $B$.  Depending on the particular method, both are in use.  Also combinations of both in decompositions like those proposed in~\cite{Liu:2018oqp} and~\cite{Rotti:2018pzi} might be of interest.  Also for this reason we show corresponding results for $Q$ and $U$ in the appendix, Figs.~\ref{fig:A3} and~\ref{fig:A4}.}  In the remainder of this paper we will investigate the $E$- and $B$-modes of full-sky polarized foreground maps.

\section{How to classify Gaussianity}
\label{sec:3}

There is a multitude of ways in which distributions can be investigated with regards to their shape.  In particular Gaussian distributions (or deviations thereof) have been in the prime focus of such methods~\citep{Dagostino:1986}.  As motivated above, we here are interested in the (dis)similarity of foregrounds with such, that follow Gaussian distributions.  While no estimator can detect \textit{any} sort of non-Gaussianity with equally high efficiency, and we do not have any preconception with regards to a specific type of distribution, we restrict ourselves to the third and fourth normalized statistical moments, the skewness and kurtosis, as done in our earlier work.

\begin{figure}
\centering
\includegraphics[width=0.4\textwidth]{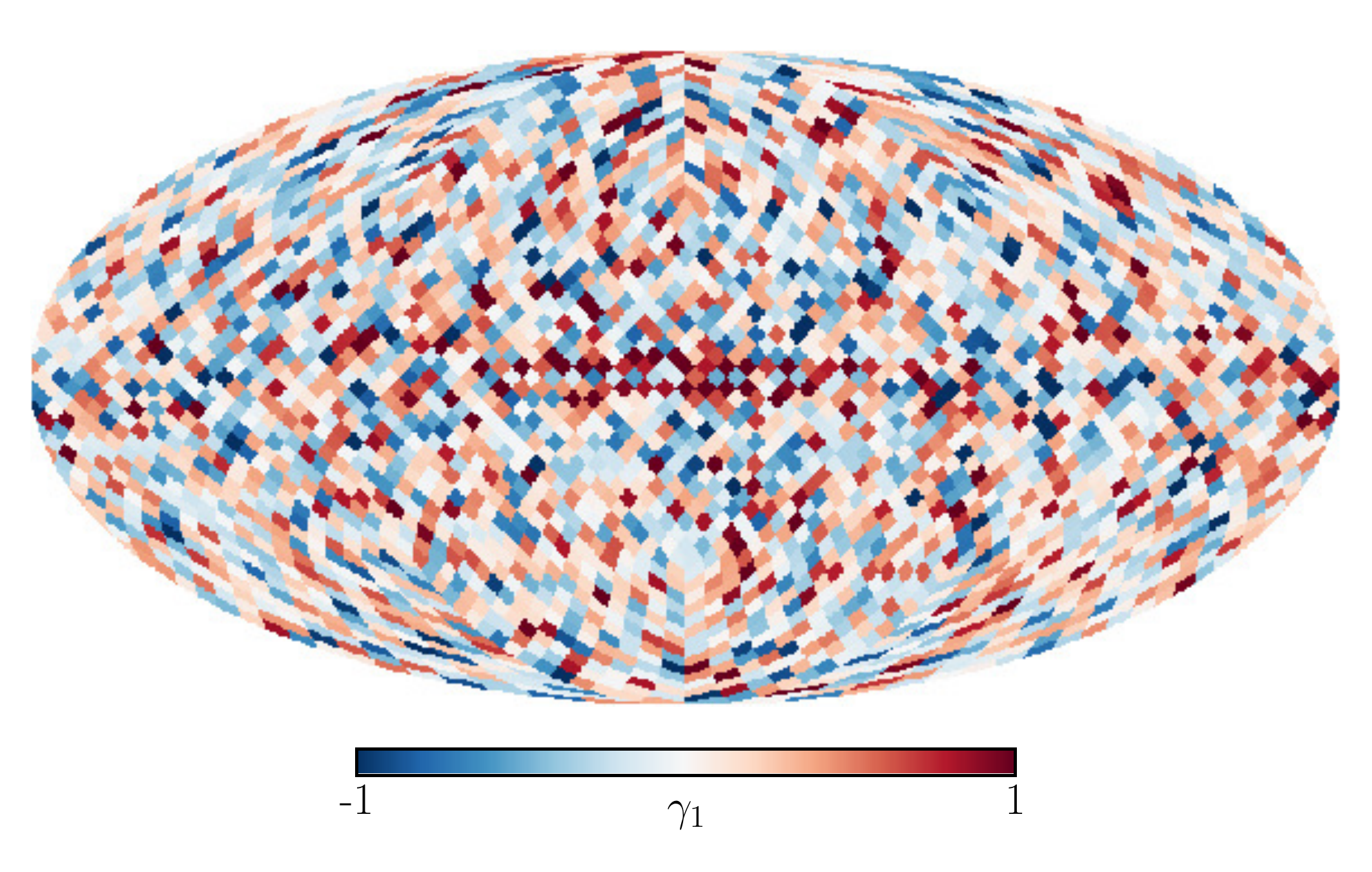}\\
\includegraphics[width=0.48\textwidth]{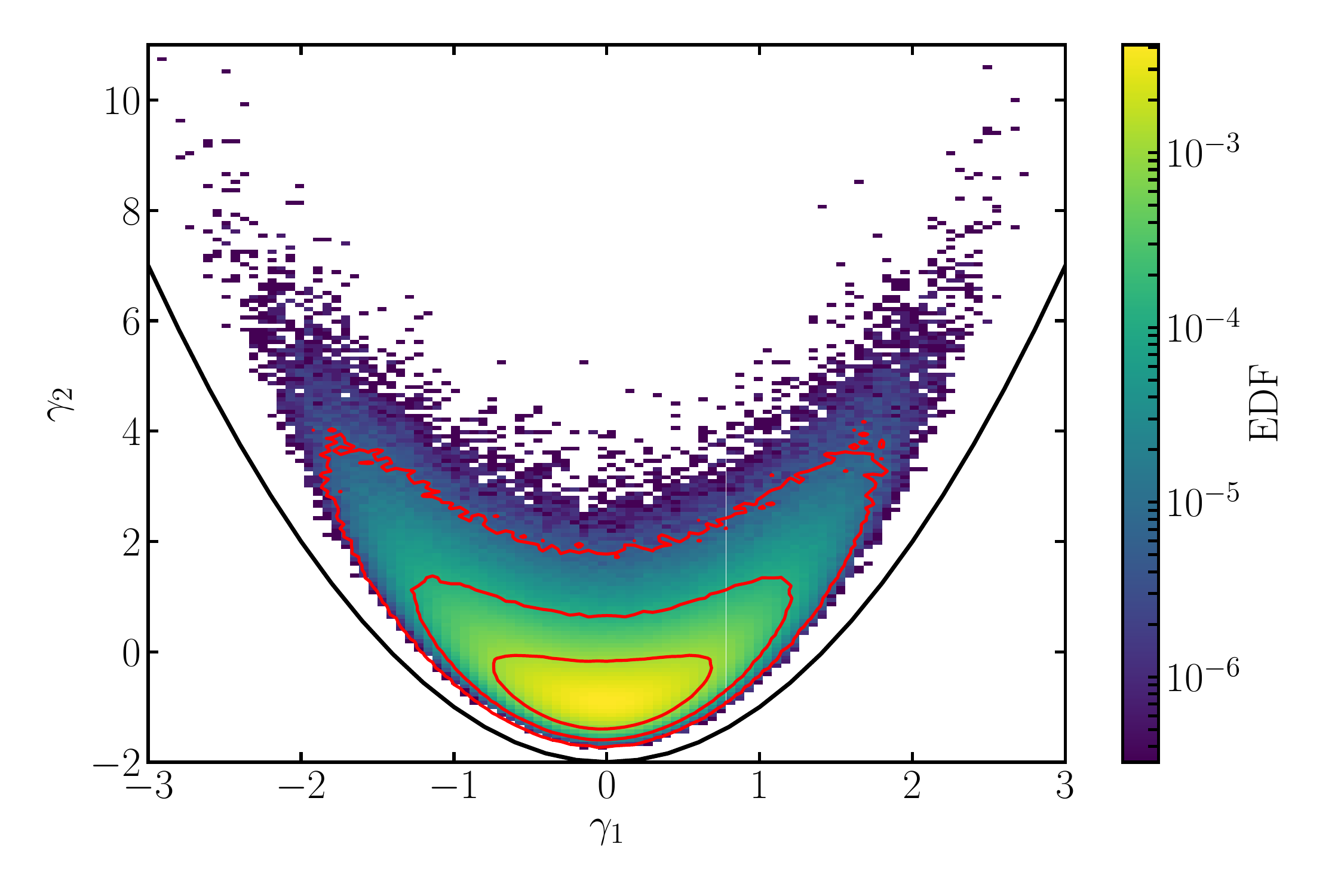}\\
\includegraphics[width=0.48\textwidth]{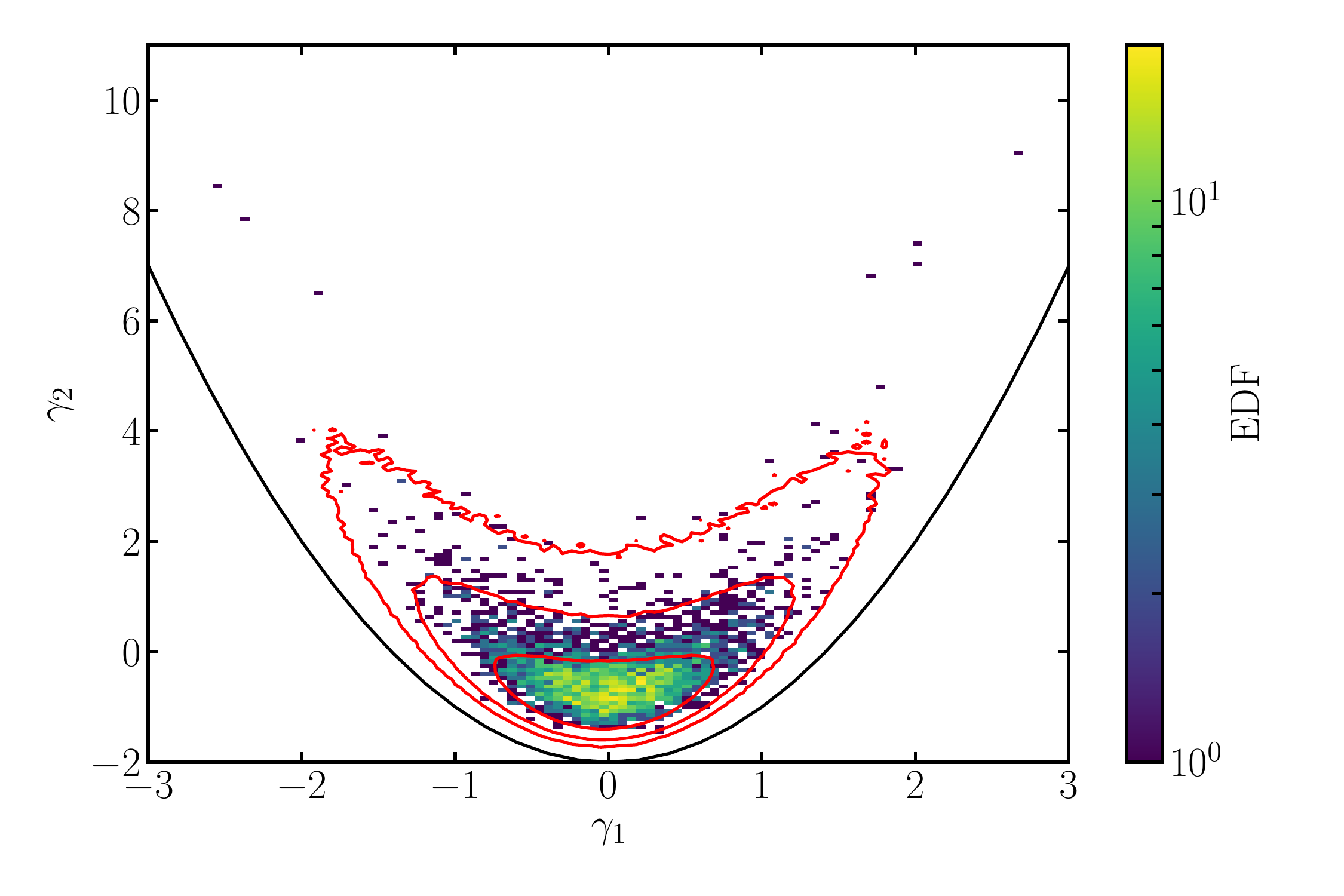}
\caption{Example demonstrating the method.  \textit{Top panel:} The skewness map of the K-band's $E$-mode.  \textit{Middle panel:} Histogram of both skewness, $\gamma_1$, and kurtosis, $\gamma_2$, from the patches of 1000 simulated sky maps based on the power spectrum of the K-band's $E$-mode.  The contours delineate 1, 2 and 3$\sigma$.  \textit{Bottom panel:} The skewness-kurtosis histogram for the K-band's $E$-mode map.  The contours from the histogram in the middle panel are used to discern Gaussian from less Gaussian patches.  }
\label{fig:1}
\end{figure}

Our aim is to test whether the emission from Galactic foregrounds on defined scales is consistent with originating from a Gaussian.  We hereto investigate the distributions of pixels within patches of that very scale as follows.  For a given map at \texttt{HEALPix}\footnote{\url{https://healpix.sourceforge.io}} resolution $\Nside=N_s$ we compute both skewness, $\gamma_1$, and excess kurtosis, $\gamma_2$, in patches defined by the pixel borders of a lower resolution, $\Nside=n_s$, as
\begin{align}
\gamma_1 = \frac{1}{N}\sum_i^N\left(\frac{x_i-m}{\sigma}\right)^3 &&
\gamma_2 = \frac{1}{N}\sum_i^N\left(\frac{x_i-m}{\sigma}\right)^4 - 3,
\end{align}
where $N=(N_s/n_s)^2$ is the number of pixels within a patch, $m$ is the mean value, and $\sigma$ is the sample standard deviation of values $x_i$ within the patch.  An example for such a map is shown in the upper panel of Fig.~\ref{fig:1}.  For a Gaussian sample we expect all odd moments to be zero and all even moments to carry no more information than already obtained from its variance.  In particular, $\langle\gamma_1\rangle=0$ and $\langle\gamma_2\rangle=-6/(N+1)$ for an uncorrelated, Gaussian sample of size $N$.  Further, the two quantities are related by the inequality~\citep{Pearson429},
\begin{align}
\gamma_2\geq\gamma_1^2-2.
\label{eq:11}
\end{align}
While similar expressions for higher moments of the expected distributions of both skewness and kurtosis can be obtained for uncorrelated draws from a Gaussian distribution, it cannot be expected that patches on a sky map contain spatially uncorrelated values.  For instance, in the case of synchrotron emission, it is the statistical properties of Galactic magnetic fields co-determining those of the emitted synchrotron signal;  a certain tendency for correlations among neighboring pixels will therefore exist.\footnote{In fact, the observed synchrotron sky arises from an interplay between many influencing factors, such as super nova rate and their spatial distribution, the cosmic ray electrons' energy and their spatial distribution, the alignment and strength of Galactic magnetic fields, absorption effects of the interstellar medium, etc.  In polarization one must further consider depolarization effects along the line-of-sight, such as the frequency dependent Faraday rotation, which depends on most of the previously mentioned quantities in a non-trivial manner.}

Spatially localized structures, i.e. correlations in the pixel domain, will generally lead to correlations in the Fourier phases, and even a Gaussian, statistically homogeneous and isotropic signal, i.e. where the phases are drawn at random from a uniform distribution, can still have `preferred clustering', as for example is expected for the CMB.\footnote{Further note that, any signal observed through an instrument is automatically convolved with the instrument's beam function, which produces named clustering to below the beam's size.}  As we wish to perform our tests in the pixel domain, we must establish a fair measure by which to estimate the degree of (or departure from) Gaussianity, given the computed values of skewness and kurtosis.  We do this by generating simulated sky maps based on the power spectrum of the investigated sky map, i.e. they all have the same power spectrum as the input map, while their phases are selected from a uniform distribution as noted above.  A set of 1000 such realizations provides us with $12000\times n_s^2$ skewness and kurtosis values which we cast into a 2-dimensional histogram (cf.~Eq.~(\ref{eq:11})), to approximate the skewness-kurtosis probability distribution from which Gaussian samples would be drawn.  Contours on this distribution will serve as guidelines when deciding on the degree of Gaussianity of the input sky map's patches, as shown in the middle and bottom panels of Fig.~\ref{fig:1}.  More detail on the computations and background information can be found in~\cite{Ben-David:2015fia}.

\section{Results}
\label{sec:4}

As described above, we are interested in extending our previous formalism of Gaussianity checks to the $E$- and $B$- polarization maps, as well as performing correlation studies on them.  The maps under investigation here are the $E$ and $B$ maps of the \textit{WMAP} K-band map~\citep{Bennett:2012zja} and the \textit{Planck} 353~GHz map~\citep{Aghanim:2018fcm} primarily, which in polarization well describe synchrotron and thermal dust polarization respectively\footnote{
We break consistency of using only \textit{Planck} maps for we want to avoid our results being contaminated by residual systematics in the \textit{Planck} 30 GHz map (see e.g.~\cite{Weiland:2018kon}, or Table~C.1 in~\cite{Liu:2018dbb}).}.  For the correlation studies we also utilize the \textit{WMAP} Ka-band polarization maps and those of the \textit{Planck} 217~GHz map.  All maps are smoothed with a $1^\circ$ beam, prior to all calculations.  The results we obtain in this manner can point out departures from Gaussian statistics of the foregrounds for the scales investigated and, in addition, locate these very departures on the sky.  Also the correlations are performed locally via mosaic correlations.

In this context it is important to emphasize that finding any one region on the $E$- or $B$-mode sky to be particularly Gaussian (or non-Gaussian) does \textit{not} equate to stating that physical processes in the same direction contribute to this finding.  This is due to the non-local definition of the $E$- and $B$-modes:  Each point on the $E/B$ sky receives contributions from all other points on the $Q/U$ sky, except its own direction~(and its antipole; for a nice explanation, see~\cite{Rotti:2018pzi}).  One might argue, that therefore the many contributions are expected to render the signal in any particular direction Gaussian by the central limit theorem.  None the less, the introductory arguments and searches for non-Gaussianity in the CMB's $E$- or $B$-modes motivate a characterization of the foregrounds along similar lines.

As in our previous work,~\cite{Ben-David:2015fia}, we present results from investigating maps at an $\Nside=512$ in patches of $\Nside=16$, which corresponds to scales of approximately $3.7^\circ$.  However, this methodology can be applied to patches of different size (and shape) for the analysis of different scales.  Despite smoothing, the presence of noise in the polarization maps of both \textit{WMAP} and \textit{Planck} influences the signal at high Galactic latitudes.  While we demonstrate in appendix~\ref{sec:B} that this does not severely affect the results in skewness and kurtosis (subsection~\ref{sec:4.1}; also Figs.~\ref{fig:A3} and~\ref{fig:A4}), we shall see the clear consequence of this in subsection~\ref{sec:4.2}.

\subsection{Skewness and Kurtosis}
\label{sec:4.1}

We calculate skewness and kurtosis values for the $E$ and the $B$ maps in \Nside=16 pixels as shown by the example in the middle panel of Fig.~\ref{fig:1}.  As described in the previous section we quantify the degree of non-Gaussianity of the respective patches on the sky as their deviations from an ensemble of Gaussian simulations.  As demonstrated in the bottom panel of Fig.~\ref{fig:1} for the $E$ map, we hereto overlay the resulting 2-dimensional skewness-kurtosis histograms of the $E$ or $B$ maps with contours from the corresponding simulations.  Using the simulations' probability distribution functions we can assign a probability to each pair of skewness/kurtosis values, and thereby to each patch.  By doing this we compute maps of departures from the Gaussian expectation in units of standard deviations for both $E$- and $B$-modes.  (In Figs.~\ref{fig:A3} and~\ref{fig:A4} we show the corresponding maps for $Q$- and $U$-modes for completeness.)

\subsubsection{Synchrotron polarization}
\label{sec:4.1.1}

\begin{figure}
    \centering
    \includegraphics[width=0.4\textwidth]{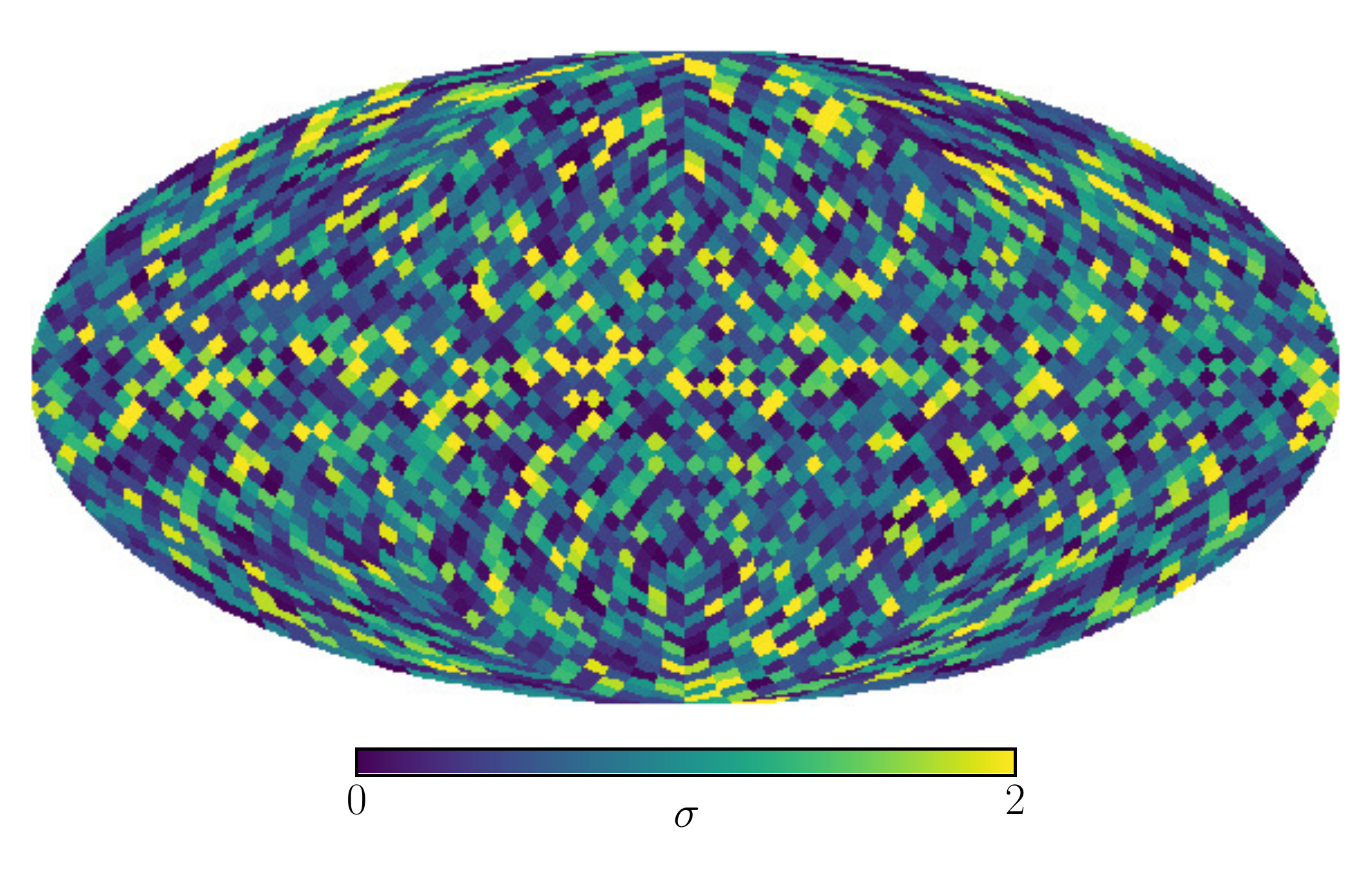}\\
    \includegraphics[width=0.4\textwidth]{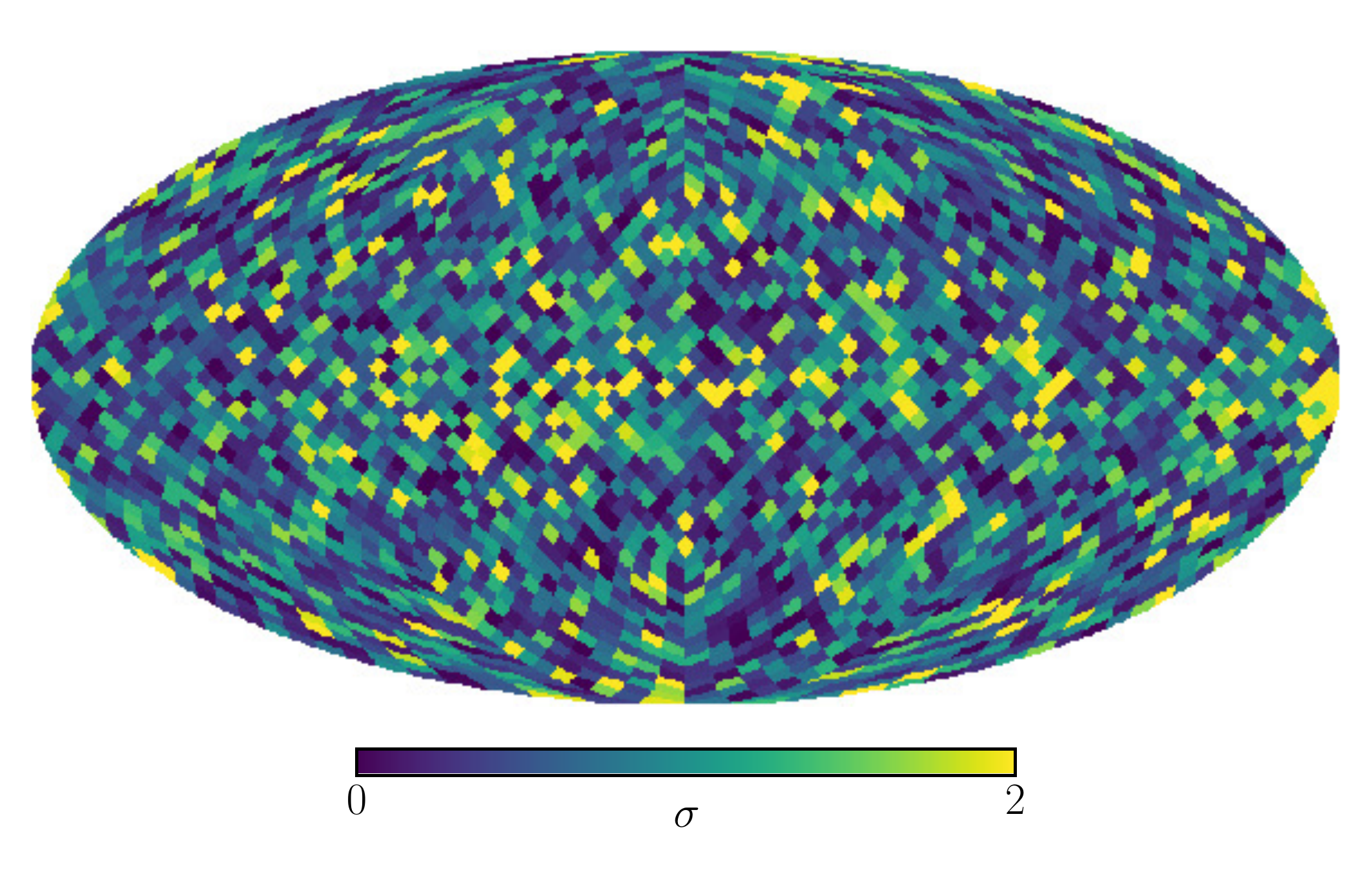}
    \caption{Deviations of the skewness and kurtosis values from their Gaussian expectation in units of standard deviations, $\sigma$, quantified by the Gaussian realizations as described in the text.  The values were computed from the \textit{WMAP} K-band $E$- (top panel) and $B$-mode (bottom panel) maps in patches of $\Nside=16$.  The standard deviations are calculated using the probability distribution functions obtained from the simulations.}
    \label{fig:2}
\end{figure}

We show maps of the standard deviations computed for the $E$- and $B$-modes of \textit{WMAP}'s K-band map in Fig.~\ref{fig:2}.  Most patches deliver values below $2\sigma$ and therefore, according to our estimator, qualify as consistent with arising from a Gaussian sample.  Apart from hints towards peculiarities along the Galactic plane the distribution of values further does not seem to prefer certain regions on the sky.  This finding resembles the one for the 408~MHz Haslam map, shown in~\cite{Ben-David:2015fia}.

\subsubsection{Dust polarization}
\label{sec:4.1.2}

\begin{figure}
    \centering
    \includegraphics[width=0.4\textwidth]{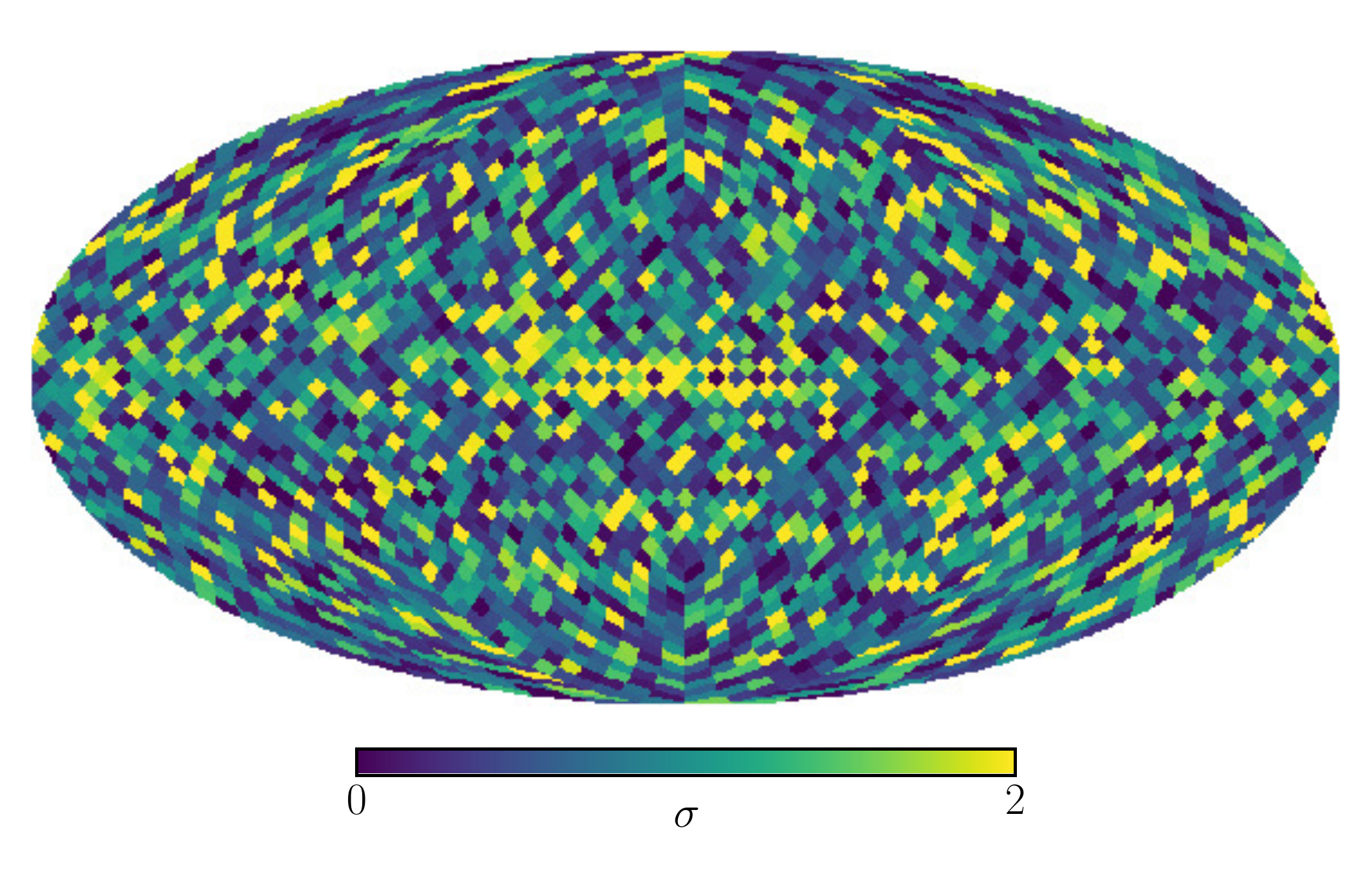}\\
    \includegraphics[width=0.4\textwidth]{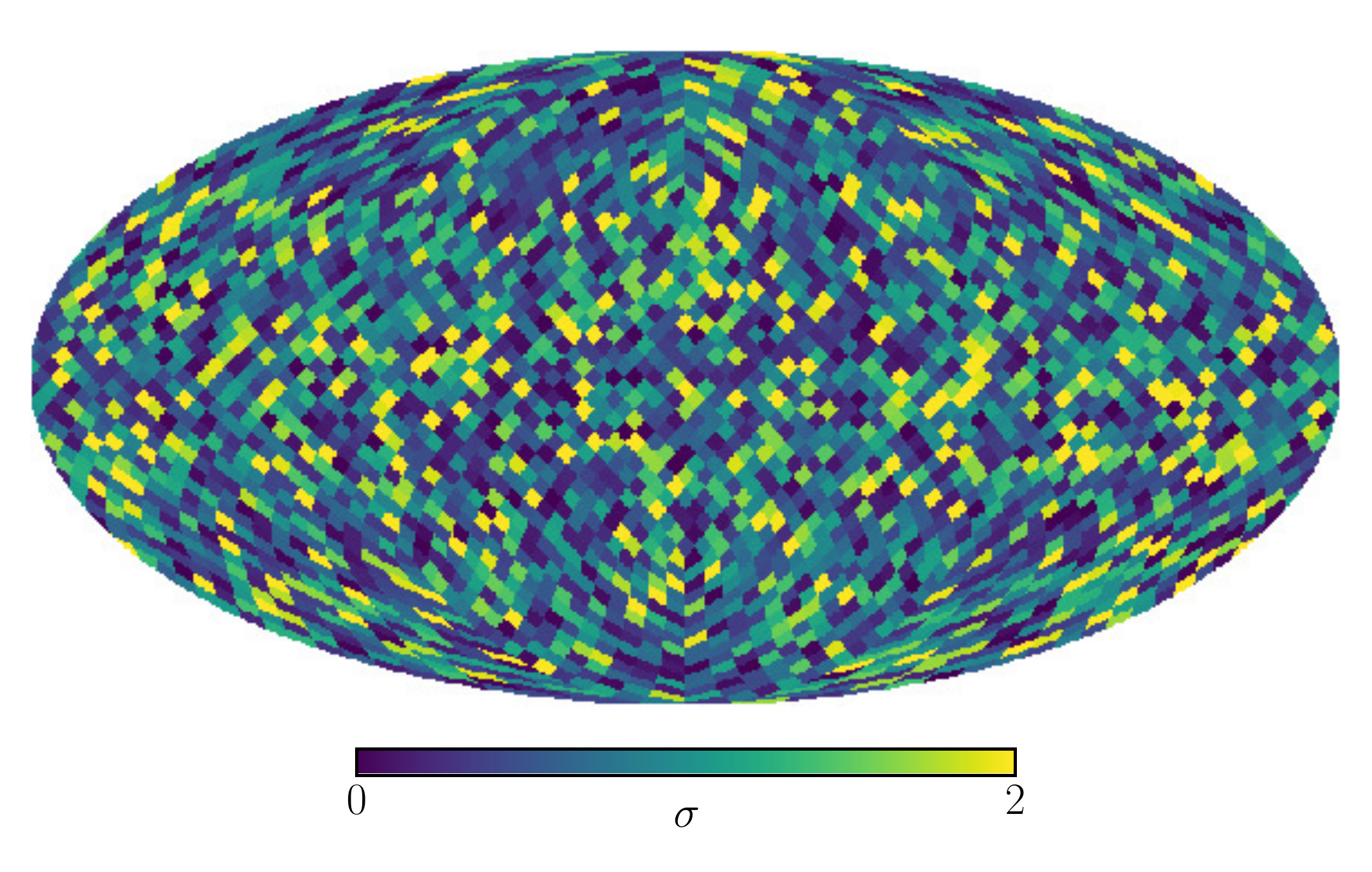}
    \caption{Same as Fig.~\ref{fig:2}, just for the polarization of the \textit{Planck} 353~GHz map.}
    \label{fig:3}
\end{figure}

Equivalent to the synchrotron results in the previous section we compute those of \textit{Planck}'s 353~GHz polarization maps for the same patch sizes.  These appear to be very similar, see Fig.~\ref{fig:3}, apart from a slight surplus of high skewness patches, which lead to a correspondingly higher abundance of $>2\sigma$ patches.  This tendency is also observed in dust \textit{temperature} maps and can be traced back to the presence of pronounced filaments and point sources in the dust maps.  While point sources are not expected to be intrinsically polarized, filaments can be visible on polarization maps.  In particular the $E$-mode will inherit power from polarized filaments, due to its sensitivity to elongated structures (see, e.g.,~\cite{Liu:2018oqp}), and indeed we find the 353~GHz $E$-mode map to give $\approx26\%$ more patches with values above $2\sigma$ than the $B$-mode map.

\subsection{Correlations}
\label{sec:4.2}

\begin{figure*}
    \centering
    \includegraphics[width=0.325\textwidth]{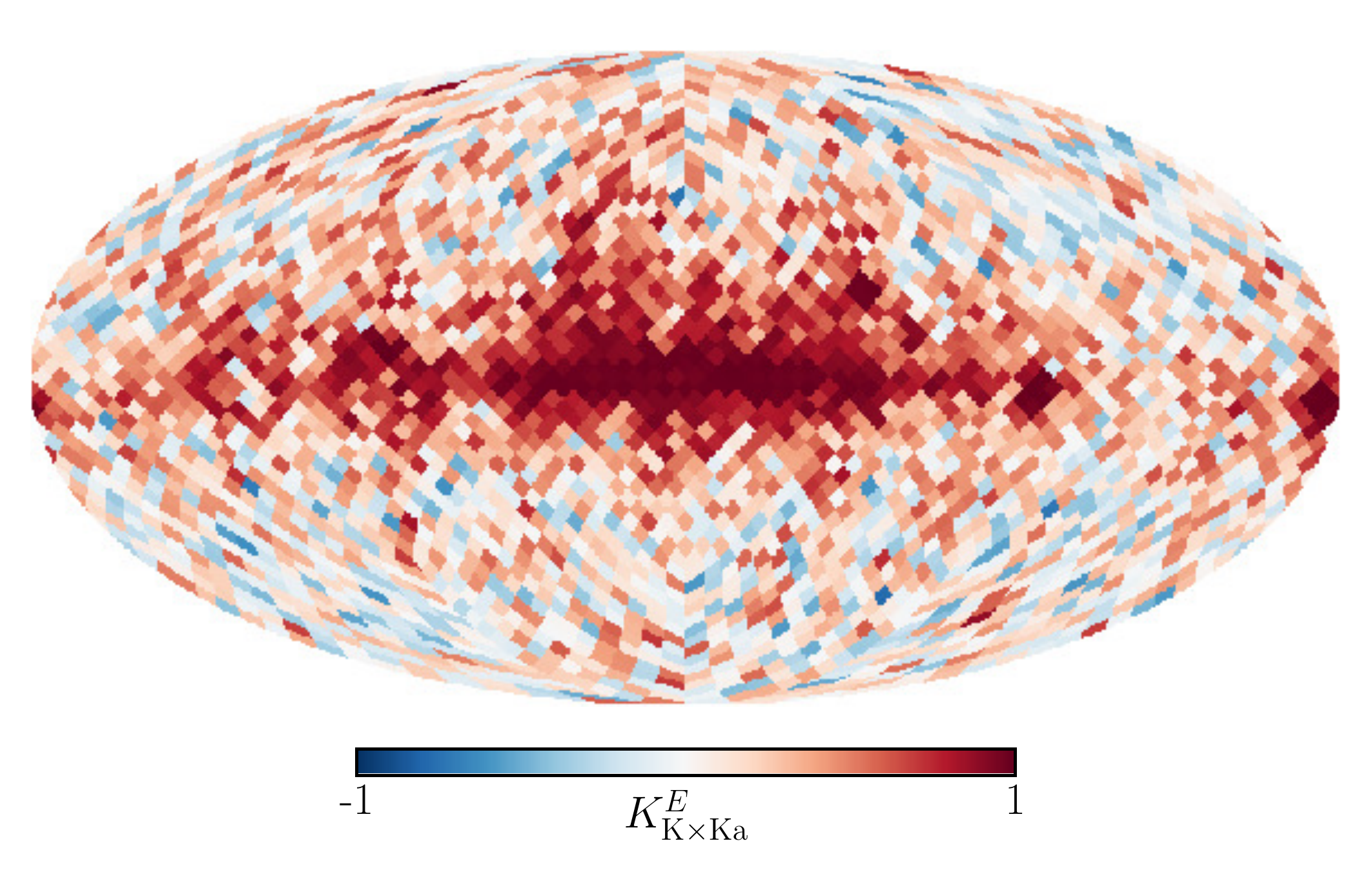}
    \includegraphics[width=0.325\textwidth]{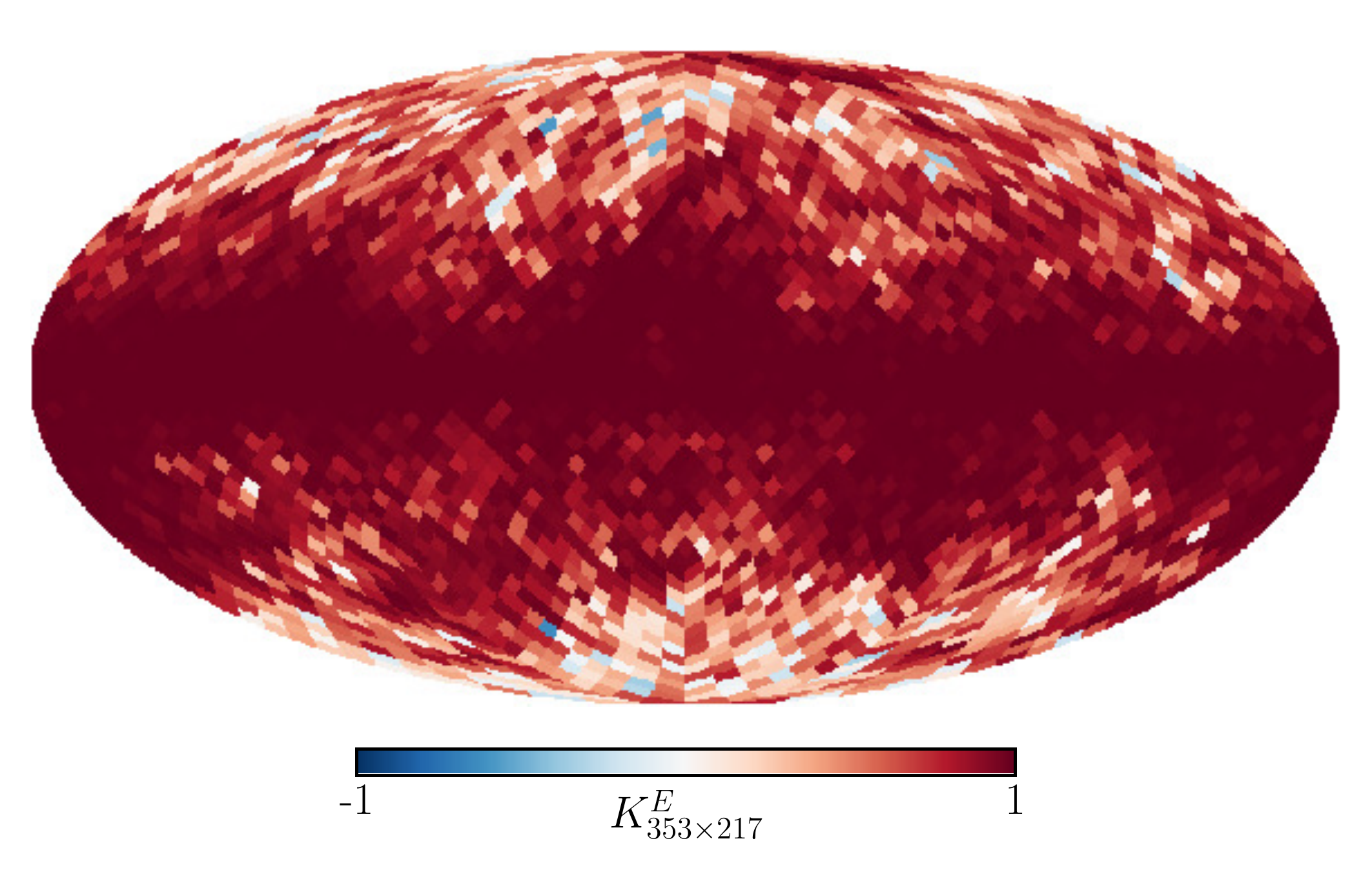}
    \includegraphics[width=0.325\textwidth]{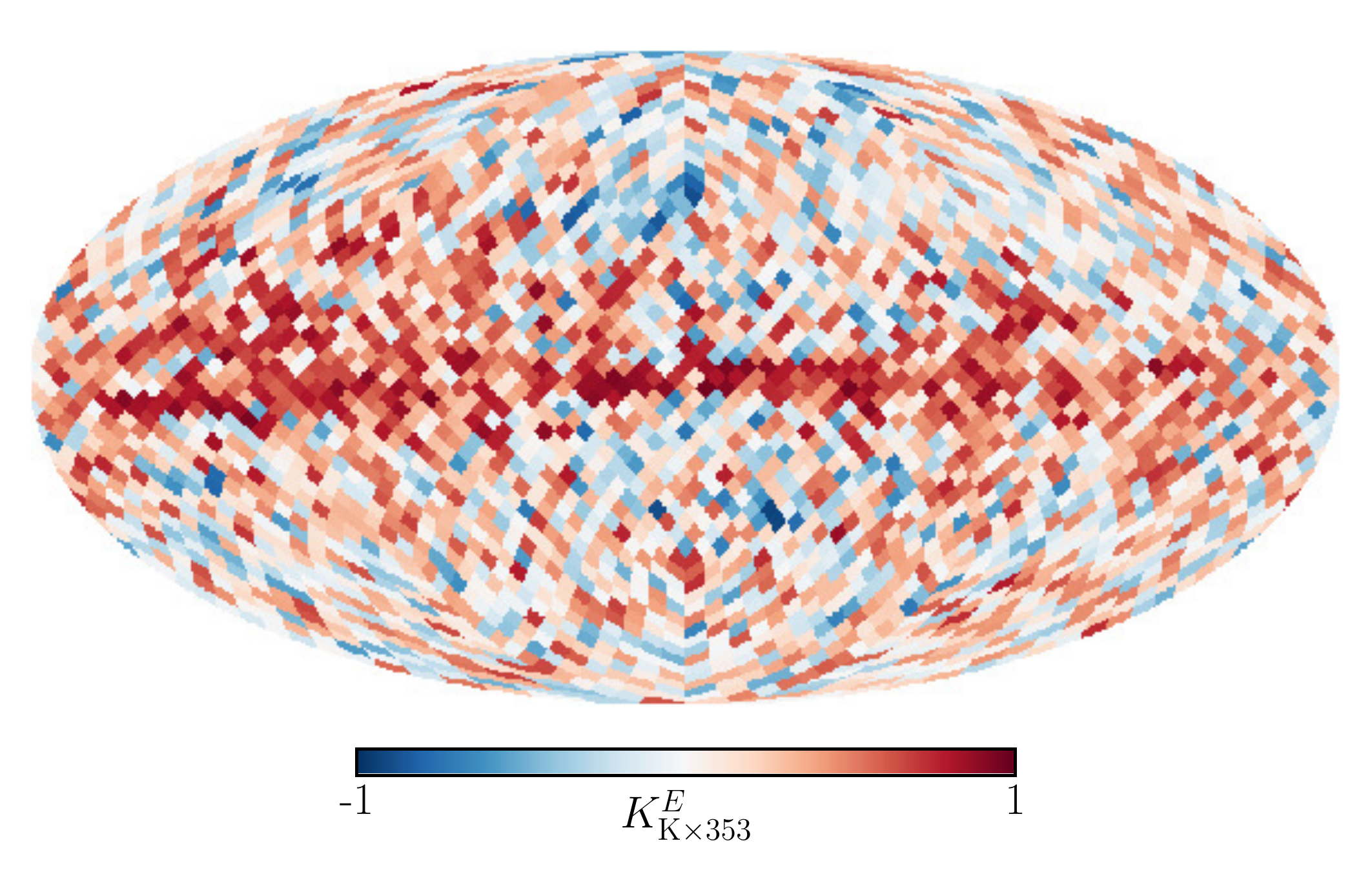}\\
    \includegraphics[width=0.325\textwidth]{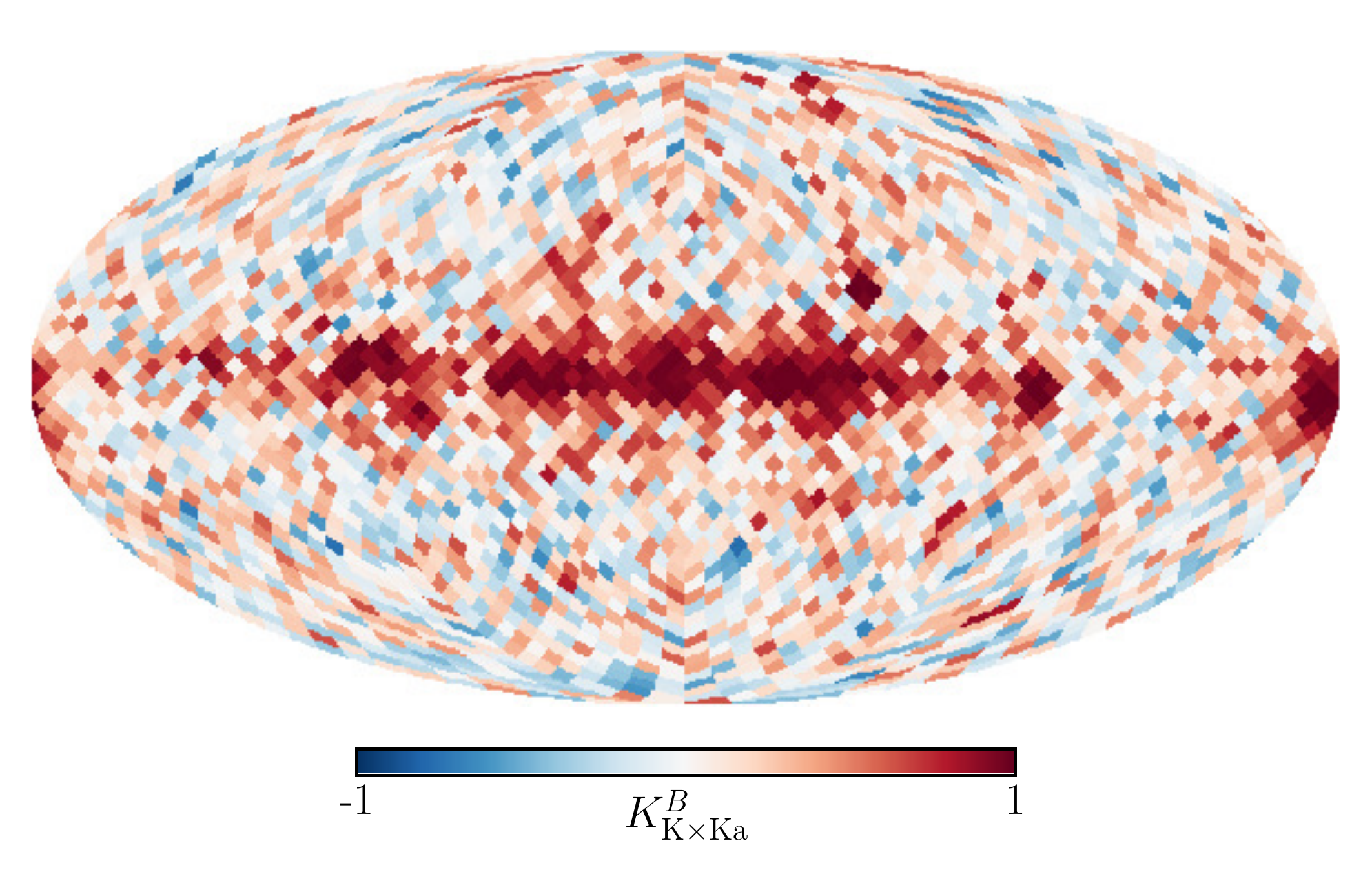}
    \includegraphics[width=0.325\textwidth]{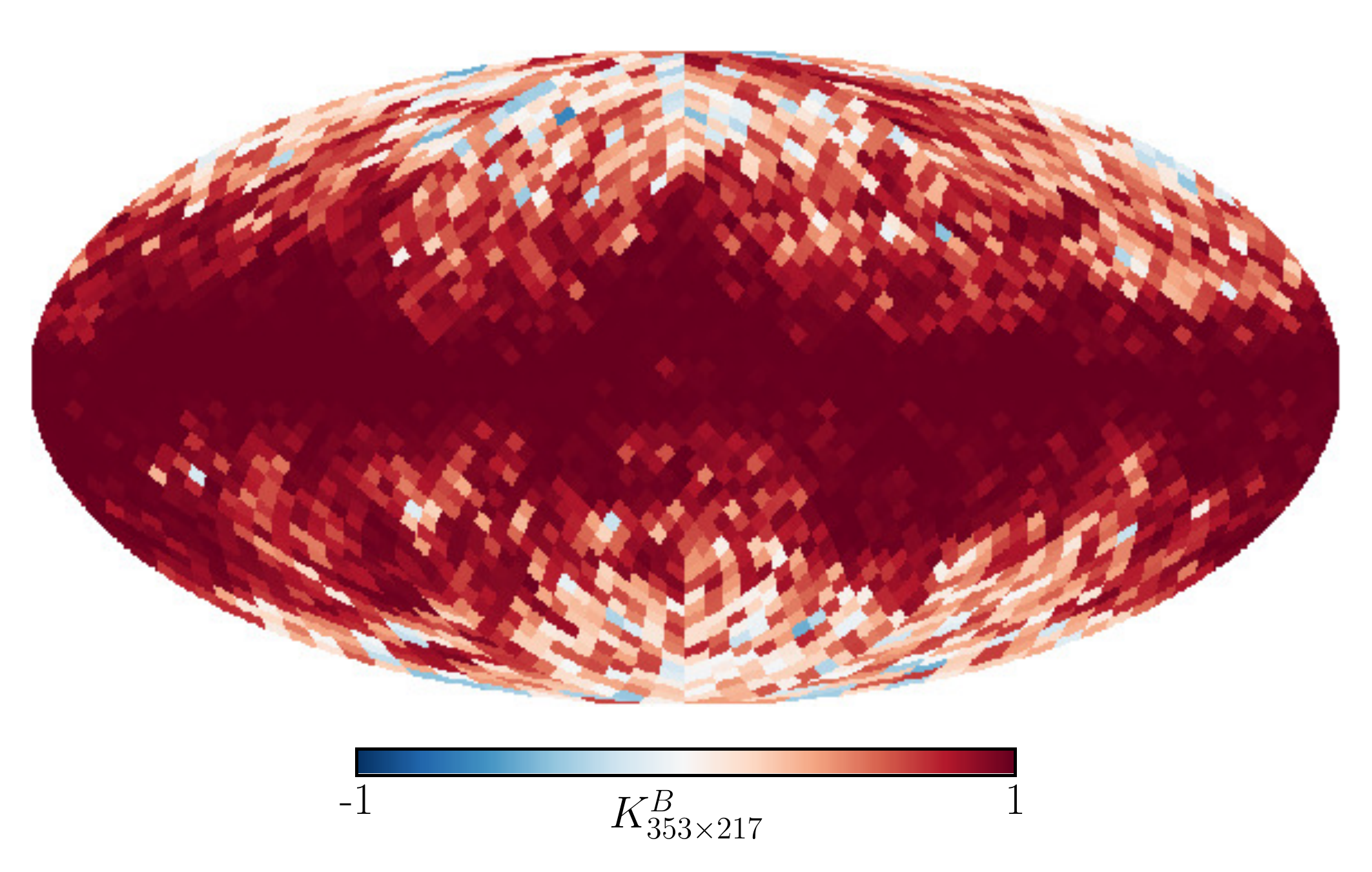}
    \includegraphics[width=0.325\textwidth]{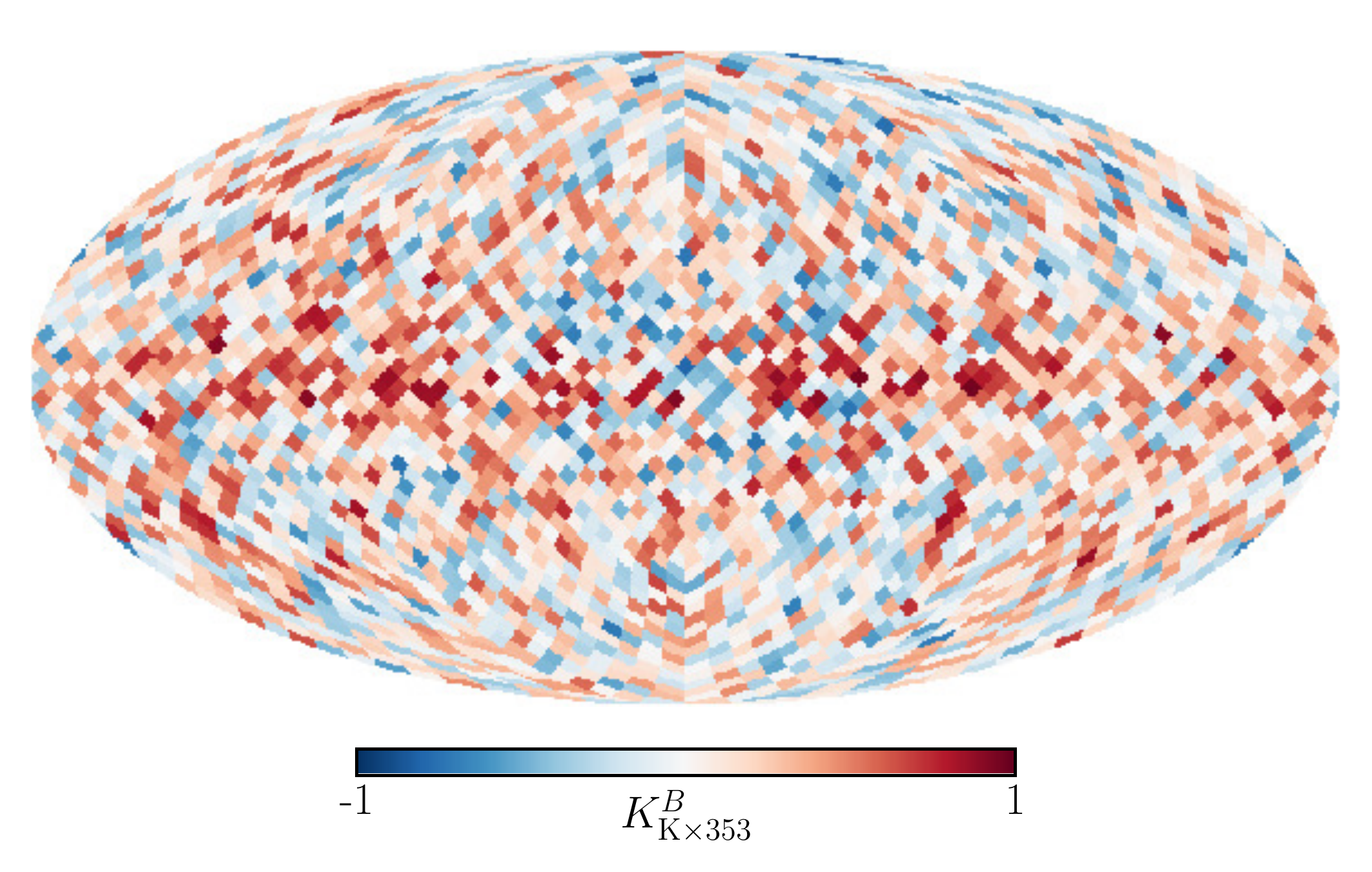}
    \caption{Mosaic correlations between $E$-mode (upper panels) and $B$-mode maps (lower panels).  \textit{Left panels:}  Correlations between \textit{WMAP} K- and Ka-band maps.  \textit{Middle panels:}  Correlations between \textit{Planck} 217 and 353~GHz maps.  \textit{Right panels:}  Correlations between \textit{WMAP} K-band and \textit{Planck} 353~GHz maps.}
    \label{fig:4}
\end{figure*}

In section~\ref{sec:2} we discussed the possibility of spatially varying spectral indices, or, more generally, spatial variation in the scaling coefficient of a `template' per foreground component.  Another point of discussion was formed by regions $\Omega$ on the sky within which foreground or noise signal are uncorrelated from band to band.  Both lead to terms in the $\mathcal{S}$ map's variance which prevent a minimization which is independent of sky location, and thereby constitute potential contamination in the final CMB product.  We shall here investigate the same patches with regards to these properties which previously were tested for Gaussianity.

As before, we at first focus on low and high frequency foregrounds separately.  We each consider correlations between the signals measured in two bands, which are dominated by the same foreground mechanism.  To make comparison to the results of the previous subsection, also here we compute Pearson correlations in patches of $\Nside=16$ -- a method known as mosaic correlation (as presented by~\cite{Verkhodanov:2009jc}, and used in~\cite{vonHausegger:2015fca} for the case of correlations among foreground temperature maps).  Any correlation short from perfect,~i.e. any change in a signal from one frequency to one nearby can only be induced by spatial variations of the spectral index (or of an equivalent scaling of the foreground), or by instrumental noise;  for the frequencies considered, no other polarized foreground components nor CMB should play a significant role besides synchrotron or thermal dust.

\subsubsection{Synchrotron polarization}
\label{sec:4.2.1}

To detect changes in the synchrotron sky in nearby frequencies we compare the \textit{WMAP} K-band (23~GHz) with the Ka-band (33~GHz) polarization maps.  Neither of free-free emission, spinning dust emission, nor molecular line emission, is expected to be polarized at a level comparable to synchrotron emission over most of the sky, such that we can assume both maps to carry signal from either synchrotron emission or noise.  We present the resulting mosaic correlation maps, $K_{\mathrm{K}\times\mathrm{Ka}}^E$ and $K_{\mathrm{K}\times\mathrm{Ka}}^B$, in the left panels of Fig.~\ref{fig:4}.  
Distinct is the area of most pronounced correlations along the Galactic plane, slightly extending also along the North Polar Spur, especially for the $E$-mode maps.  While positive correlations overweigh, at intermediate to high Galactic latitudes the correlations weaken and expose many patches with correlations close to zero, see also Fig.~\ref{fig:A1}.  In order to understand the origin of this reduced correlation we investigated \textit{WMAP} K- and Ka-band single-year maps which showed that the dominating contributor at high Galactic latitudes is most likely noise.  However, recall that it is irrelevant to the point presented in section~\ref{sec:2} whether this loss in correlation arises from changes in the foregrounds' morphology or from uncorrelated instrumental noise in the two bands -- both present challenges for ILC-like methods.

\subsubsection{Dust polarization}
\label{sec:4.2.2}

Applying the same method to the 217~GHz and 353~GHz \textit{Planck} maps to obtain results for the polarized thermal dust sky, leads to the correlations presented in the middle panels of Fig.~\ref{fig:4}.  Similar to the case at low frequencies high correlation can be observed close to the Galactic plane and lower to vanishing correlation towards high Galactic latitudes.  The larger abundance of high correlation patches, see also Fig.~\ref{fig:A2}, is largely due to higher signal-to-noise ratios in the maps (a conclusion also supported by analyzing half-mission maps), but could also indicate generally lower amount of spectral index variation in thermal dust emission.

\subsubsection{Correlations between synchrotron and dust}
\label{sec:4.2.3}

Before concluding we still need to consider correlations between synchrotron and dust emission.  In the case of more than a single foreground the variance $\langle\mathcal{S}^2\rangle_\Omega$  will also contain cross-terms between the different foreground components such that the weights $w_\nu$ need to be determined according to high or low cross-correlation.  We show the corresponding mosaic correlation maps in the right panels of Fig.~\ref{fig:4} for both $E$- and $B$-modes of the \textit{WMAP} K-band and the \textit{Planck} 353~GHz map.  The tendency in both is higher correlations along the Galactic plane.  However, away from the Galactic plane no strong correlation between the polarized emission of synchrotron and thermal dust emission is established on the scales given by the patch size investigated here.  We further point out that while the high Galactic latitudes are most likely, as before, affected by instrumental noise, noticeable differences between the correlations at low Galactic latitudes from $E$- and $B$-modes can be seen -- those latitudes where instrumental noise is subdominant.  We return to this point in the discussion.  However, higher sensitivity observations will be needed to characterize the foregrounds' polarized emission better both at low and at high frequencies.

\section{Discussion and conclusion}
\label{sec:5}

While also of astrophysical interest statistical investigations of Galactic foregrounds also have great relevance in regards to sufficient foreground removal, as we showed in this paper.  In particular distinct statistical properties of $E$- and $B$-mode signals require an adequate, perhaps separate treatment in foreground separation algorithms.

In this work we investigated statistics of the two strongest CMB foregrounds in polarization, synchrotron radiation and thermal dust emission, in order to draw conclusions on the feasibility of obtaining a clean CMB map.  In particular variations of foreground properties across the sky were of motivated interest wherefore we employed two methods which both work in predefined patches on the sky -- the skewness-kurtosis method for classification of foreground maps as Gaussian processes~\citep{Ben-David:2015fia}, and the mosaic correlations~\citep{vonHausegger:2015fca}.  The $E$- and $B$-mode maps under investigation were smoothed to $1^\circ$ and the selected patches were of extent $\sim3.7^\circ$, which can be roughly translated into the multipole range $\ell\in[50,180]$.  We summarize our findings as follows:
\begin{itemize}
\item
On scales between approximately $1^\circ$ and $3.7^\circ$ the $E$- and $B$-mode maps of both synchrotron and thermal dust polarization exhibit distributions consistent with those expected from Gaussian realizations over most of the sky, with preferred regions of departure only along the Galactic plane, cf.~Figs.~\ref{fig:2} and~\ref{fig:3}.  
\end{itemize}
This might prove itself helpful in the construction of polarized foreground simulations \`a la~\cite{Hervias-Caimapo:2016crc}.  Also studies on measuring polarization of the 21-cm line~\citep{Babich:2005sb} require simulations of polarized radio foregrounds.  Our findings can be seen as an addition to implementing or justifying assumptions about Gaussianity of foregrounds in such simulations, also in polarization~(e.g.~\cite{Jelic:2008jg}).\footnote{Since our previous findings~\citep{Ben-David:2015fia} for a temperature map of synchrotron emission could be confirmed by~\cite{Rana:2018oft}, it would be interesting to see whether the presented results are also confirmed using equivalent methods for polarization such as~\cite{Chingangbam:2017uqv}.}

\begin{itemize}
\item
On the same scales we find spatial variation of the frequency spectra of both synchrotron and thermal dust polarization to be negligible along the Galactic plane, cf.~Fig.~\ref{fig:4}.
\item
At intermediate and higher Galactic latitudes increased instrumental noise prevents us from drawing conclusions about the spectral properties.  However, the observed decrease in correlations also in the case of noise will impede ILC-like foreground removal algorithms, as elaborated in section~\ref{sec:2}.
\item
Mosaic correlations between synchrotron and dust polarization maps were distinctly different for $E$- and $B$-mode maps along the Galactic plane.
\end{itemize}
Given this last point, weights determined from the ILC approach would therefore be necessarily different for a foreground separation in $E$- or $B$-modes.  Separation of foreground and CMB performed, for example, with the Stokes parameter $Q$ and $U$ maps, would thereby mix the distinct statistical properties of $E$- and $B$-mode signals.  Exploration of new methods of foreground separation for polarized signals might therefore be desirable.  Furthermore, depolarization effects add additional uncertainty and/or bias to foreground cleaning algorithms.  Statistical investigations as those presented here offer means to understand such effects or their influences on the final CMB product in more detail, especially once higher fidelity polarization data become available.

It should not be omitted to mention that correlations between synchrotron and dust polarization have been subject to a many of recent studies to assess the potential level of foreground contribution at those frequencies where the CMB signal is strongest~\citep{Choi:2015xha,Krachmalnicoff:2015xkg,Krachmalnicoff:2018imw,Akrami:2018wkt}.  Their findings underline the existence of these correlations among the components also at high Galactic latitudes, and correlations were quantified for different scales by computing the corresponding cross-power spectra.  However, they do not consider spatial variation of the correlation other than imposing Galactic masks of different extent.  For the scales chosen in this work, we pointed out these spatial variations here, and, in addition to the differences between those from $E$- and $B$-modes, showed their relation to principles in current foreground separation~algorithms.\\

In light of the increasingly precise measurements at microwave frequencies performed for the observation of the CMB, an equally precise understanding of the behavior of those sources interfering with a clean measurement is required.  Given that foreground separation algorithms and the subsequent inclusion of the CMB products in a combined framework for determining the cosmological parameters focus on the less contaminated regions of the sky, away from the Galactic plane, these results will need to be taken into account in further analyses of the results.  In other words, the supposedly cleanest regions of the sky might contain the most stubborn foregrounds.

\section*{Acknowledgements}

We thank Pavel Naselsky for helpful and exciting discussions.  This work was funded in part by the Danish National Research Foundation (DNRF) and by Villum Fonden through the Deep Space project.  Hao Liu is supported by the Youth Innovation Promotion Association, CAS.  Some of the results in this paper have been derived using the \texttt{HEALPix}~\citep{Gorski:2004by} package implemented into \texttt{healpy}.  We thank both the referee and the editor for their constructive comments.

%%%%%%%%%%%%%%%%%%%%%%%%%%%%%%%%%%%%%%%%%%%%%%%%%%

%%%%%%%%%%%%%%%%%%%% REFERENCES %%%%%%%%%%%%%%%%%%

\bibliographystyle{mnras}
\bibliography{ref.bib}

%%%%%%%%%%%%%%%%%%%%%%%%%%%%%%%%%%%%%%%%%%%%%%%%%%

%%%%%%%%%%%%%%%%% APPENDICES %%%%%%%%%%%%%%%%%%%%%

\appendix

\section{Additional figures}
\label{sec:A}

\begin{figure}
    \centering
    \includegraphics[width=0.43\textwidth]{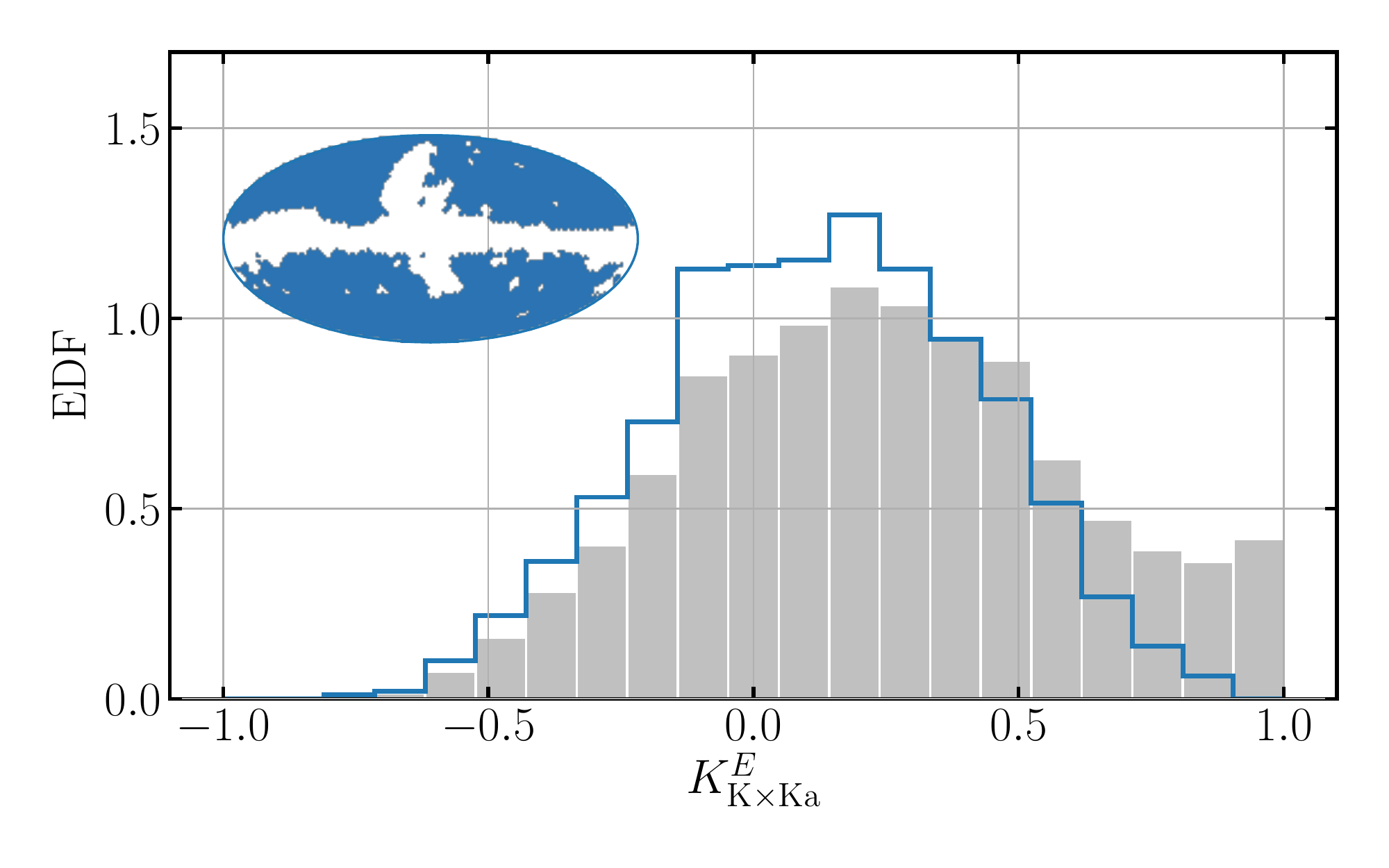}\\
    \includegraphics[width=0.43\textwidth]{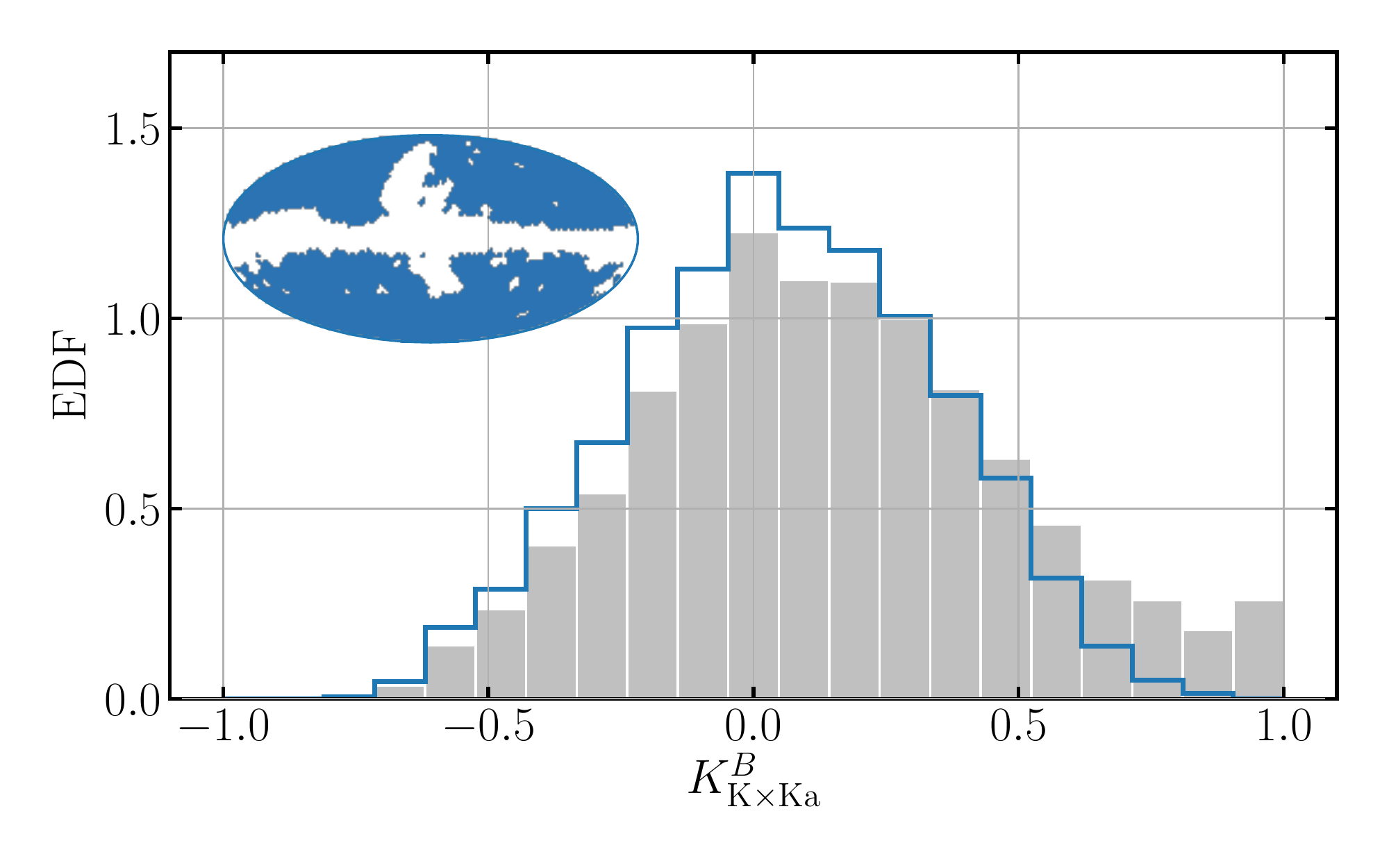}
    \caption{Histograms of the mosaic correlations between \textit{WMAP} K- and Ka-band $E$-mode (upper panel) and $B$-mode maps (lower panel), as shown in Fig.~\ref{fig:4}.  The inset shows in blue those regions from which the blue distribution is plotted, corresponding to the \textit{WMAP} polarization mask.  The gray histogram contains the values from the full sky.}
    \label{fig:A1}
\end{figure}

\begin{figure}
    \centering
    \includegraphics[width=0.43\textwidth]{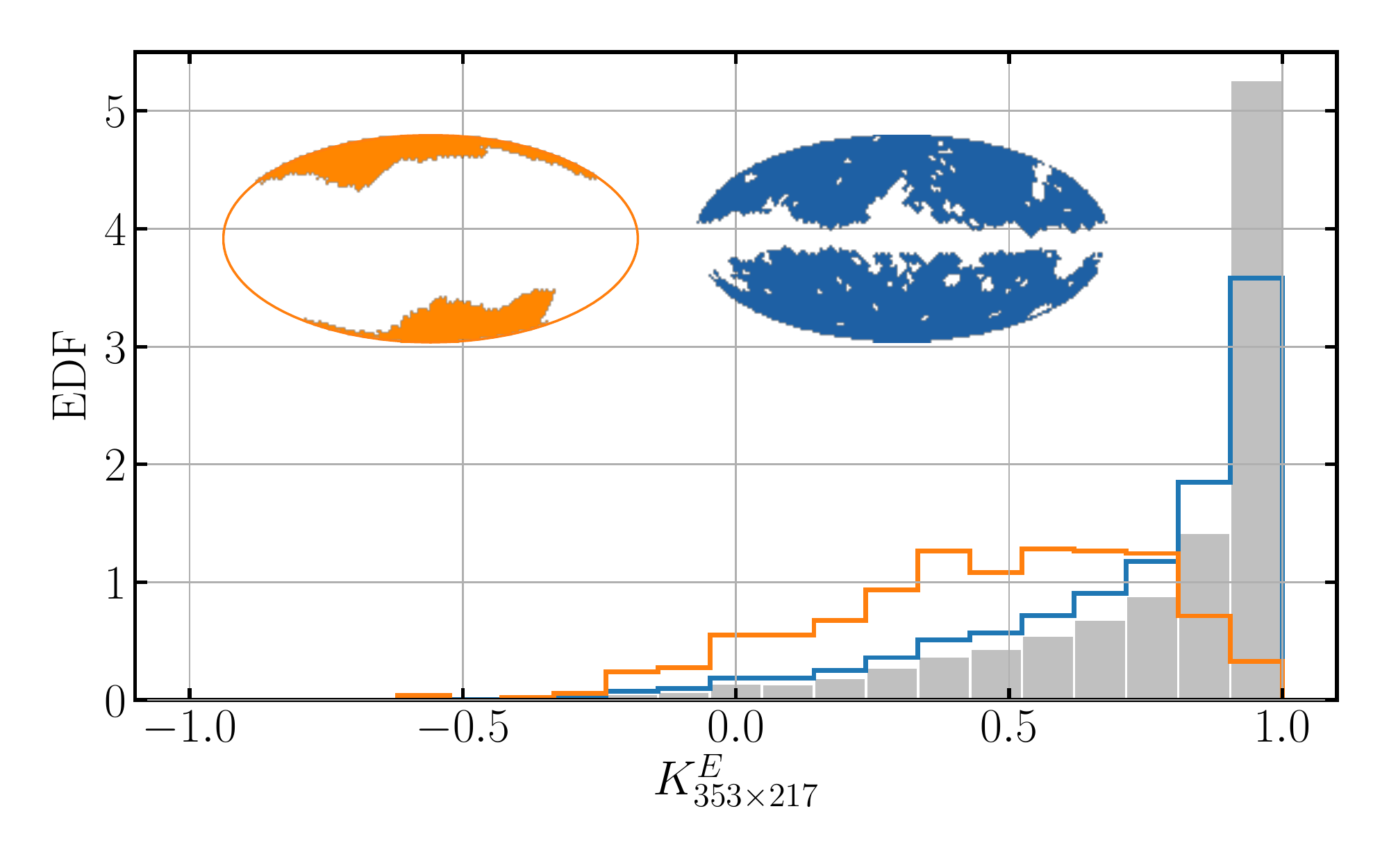}\\
    \includegraphics[width=0.43\textwidth]{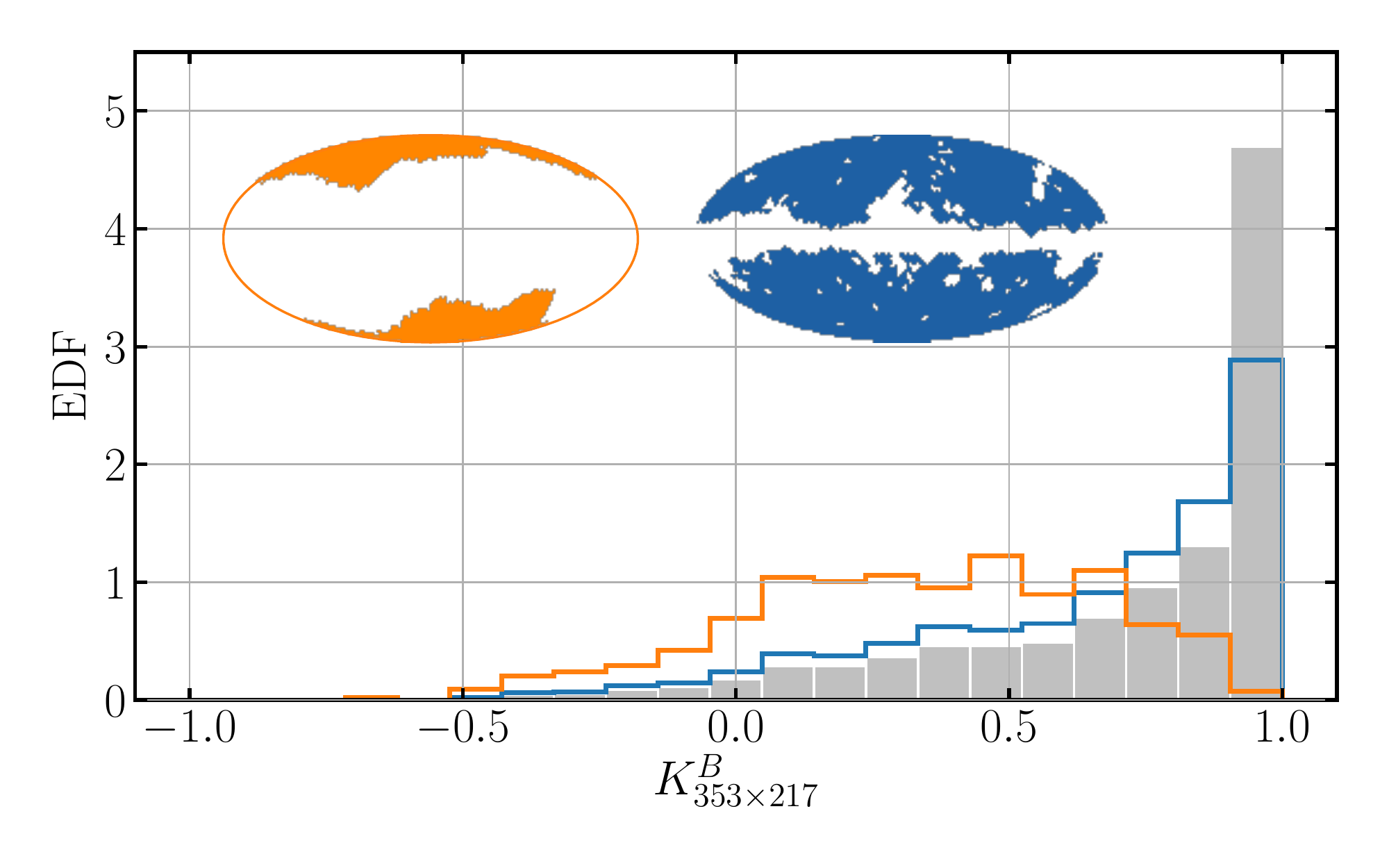}
    \caption{Same as Fig.~\ref{fig:A1}, but for mosaic correlations between the \textit{Planck} 353~GHz and the 217~GHz maps.  The blue mask here is the \textit{Planck} polarization mask.  In addition we select patches via a more aggressive mask, the \textit{Planck} Gal20 mask, shown in orange.}
    \label{fig:A2}
\end{figure}

\begin{figure}
    \centering
    \includegraphics[width=0.4\textwidth]{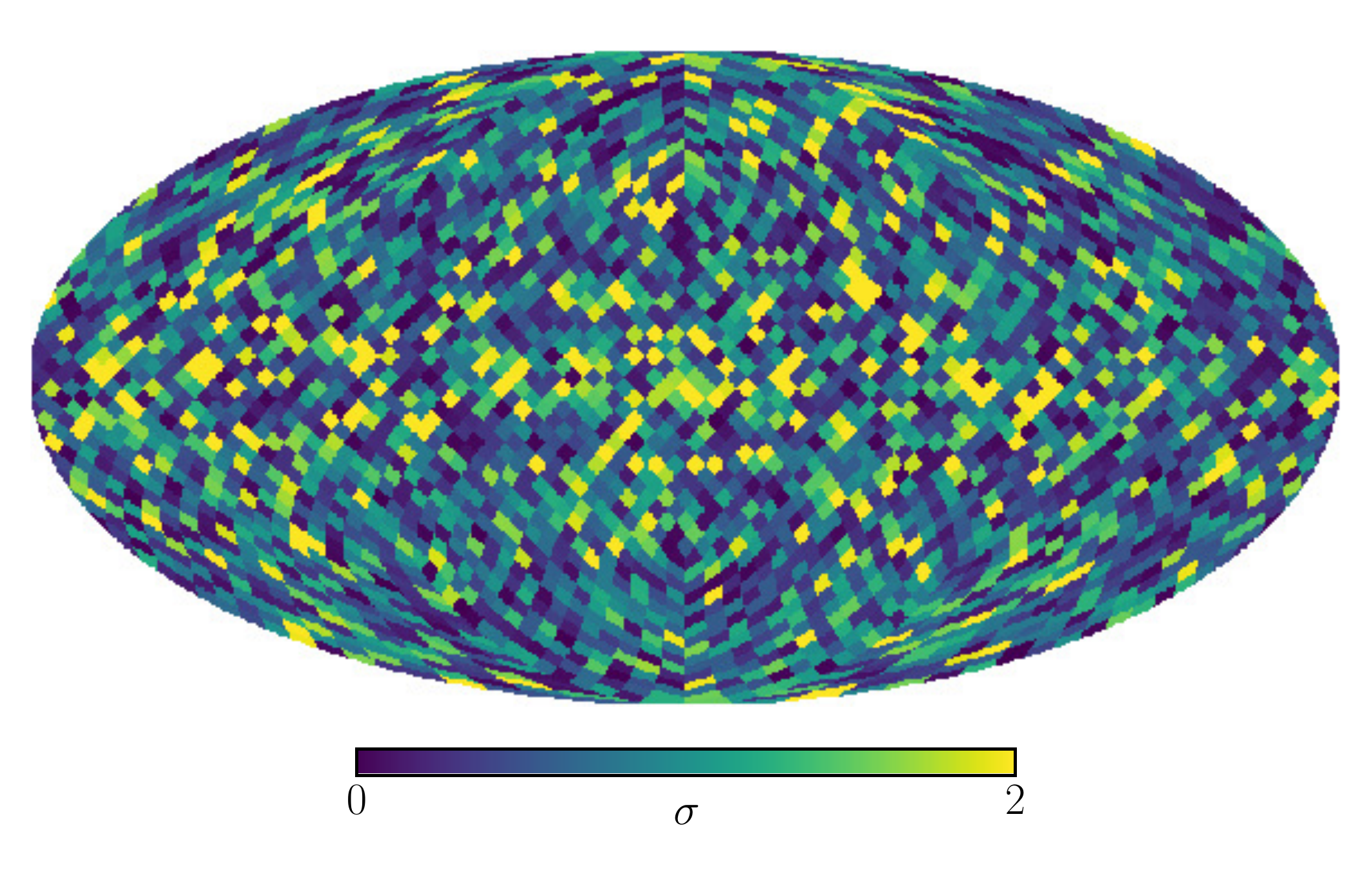}\\
    \includegraphics[width=0.4\textwidth]{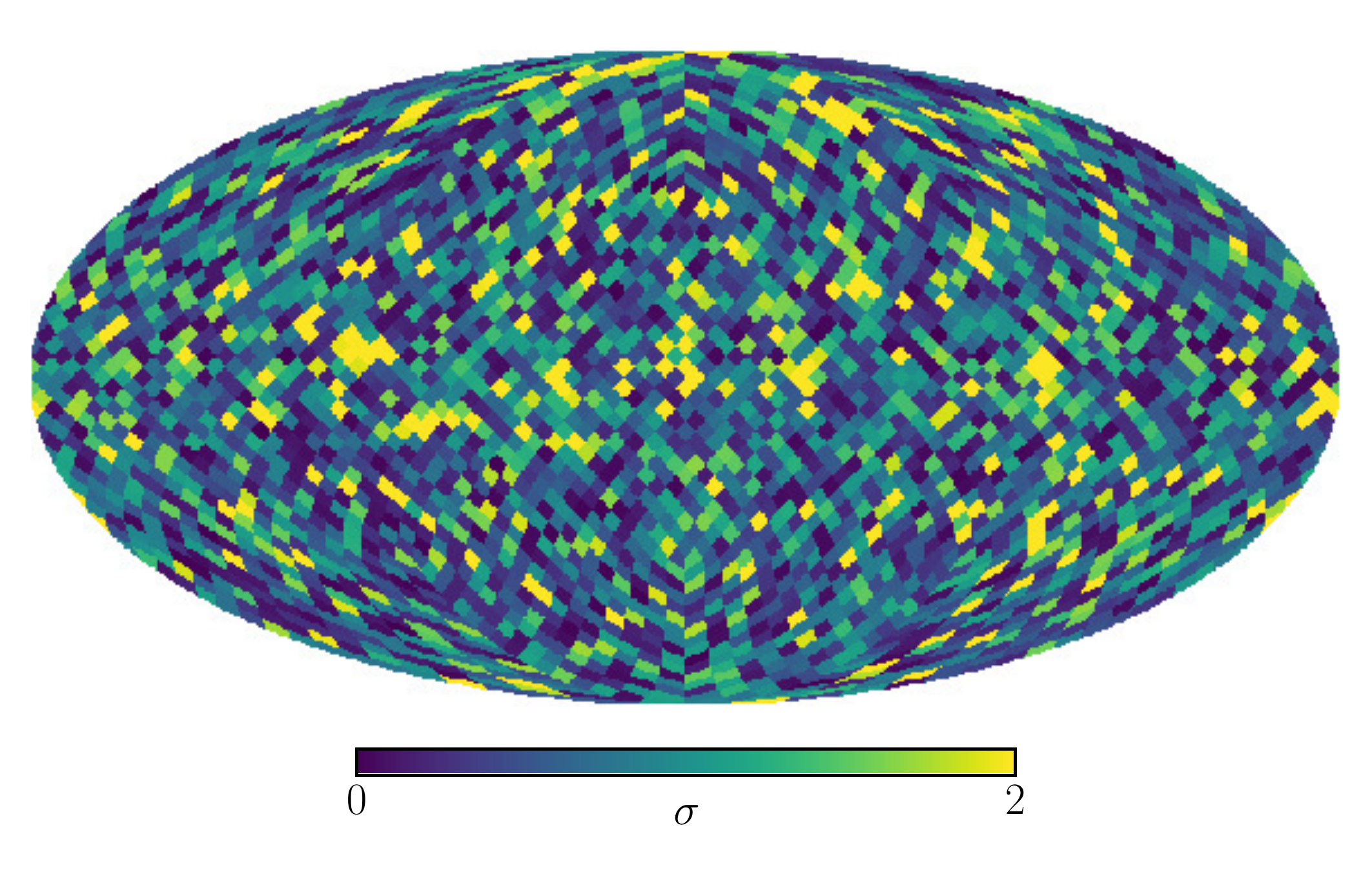}
    \caption{Deviations of the skewness and kurtosis values from their Gaussian expectation in units of standard deviations, $\sigma$, quantified by the Gaussian realizations as described in the text.  The values were computed from the \textit{WMAP} K-band $Q$- (top panel) and $U$-mode (bottom panel) maps in patches of $\Nside=16$.  The standard deviations are calculated using the probability distribution functions obtained from the simulations.}
    \label{fig:A3}
\end{figure}

\begin{figure}
    \centering
    \includegraphics[width=0.4\textwidth]{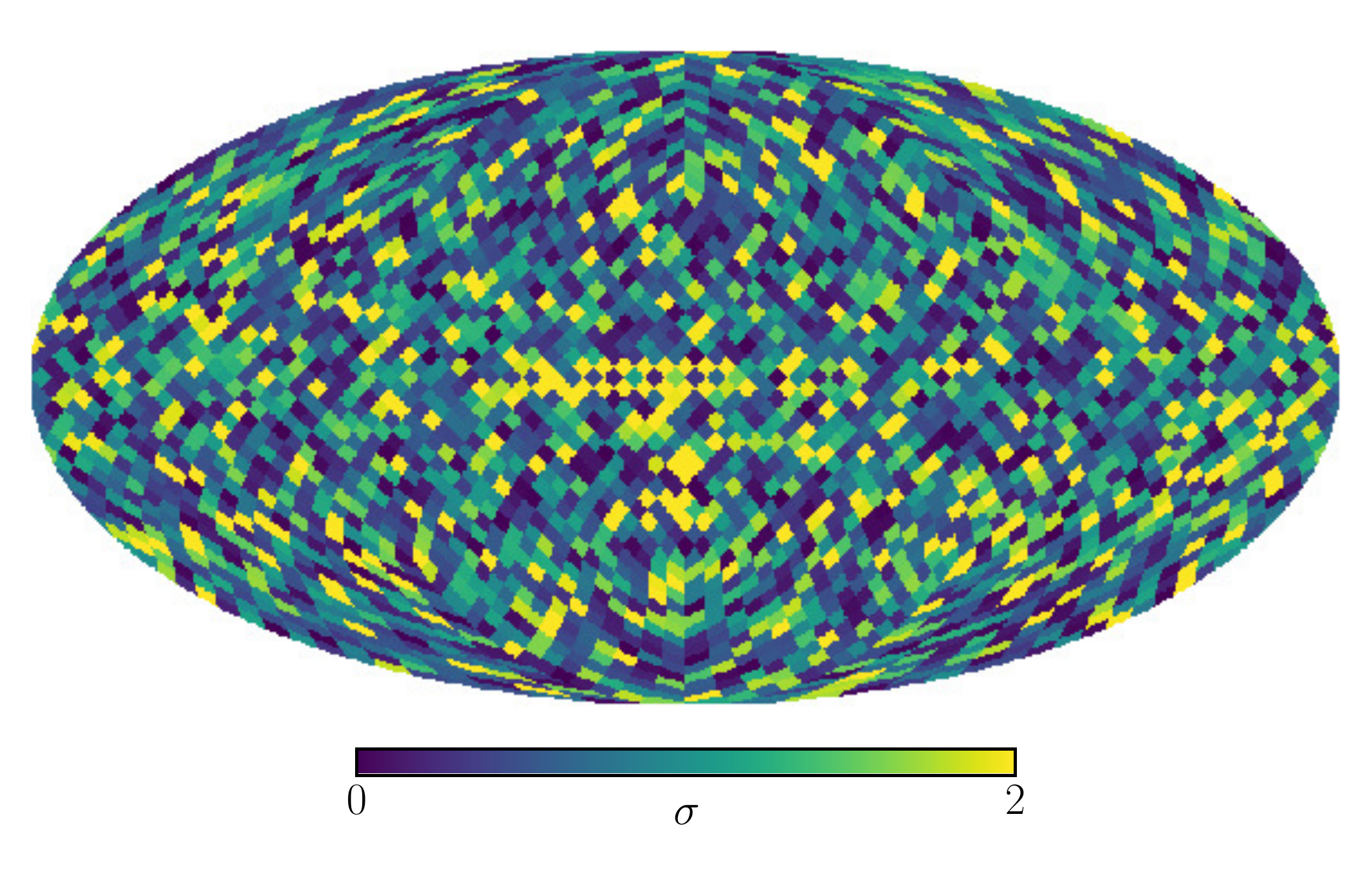}\\
    \includegraphics[width=0.4\textwidth]{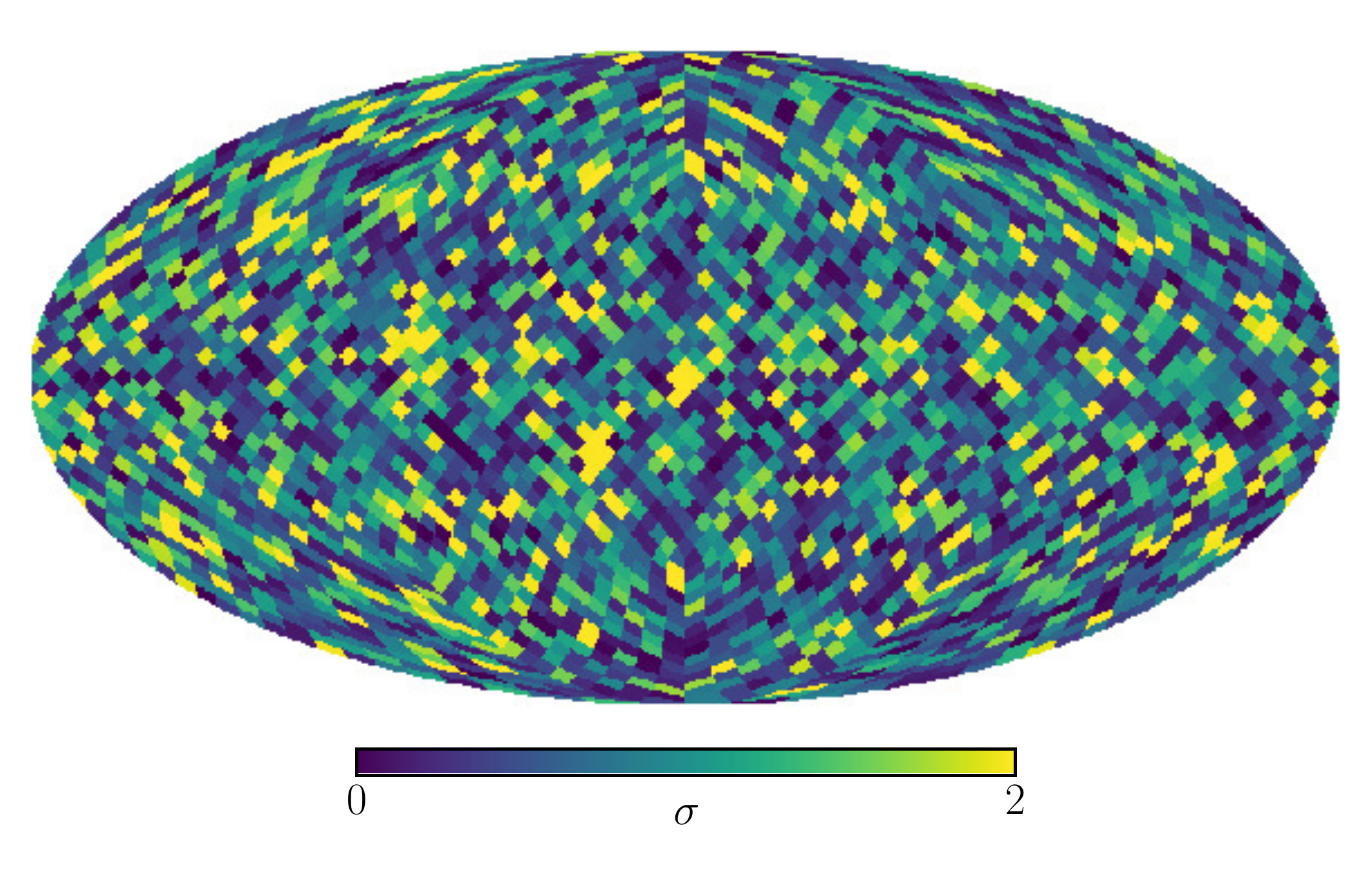}
    \caption{Same as Fig.~\ref{fig:A3}, just for the polarization of the \textit{Planck} 353~GHz map.}
    \label{fig:A4}
\end{figure}

\newpage

\section{Weighted moments}
\label{sec:B}

\begin{figure}
    \centering
    \includegraphics[width=0.4\textwidth]{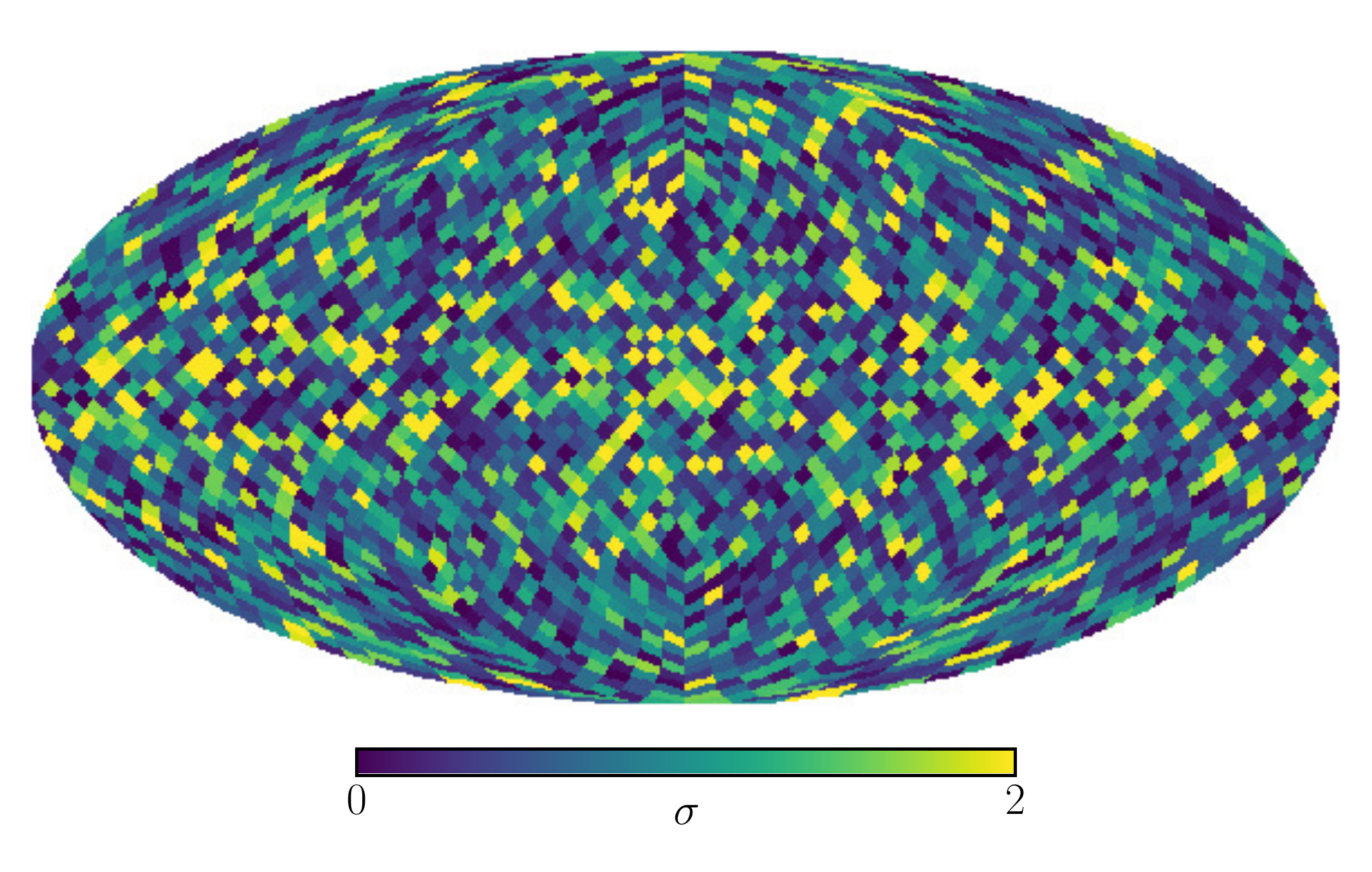}\\
    \includegraphics[width=0.4\textwidth]{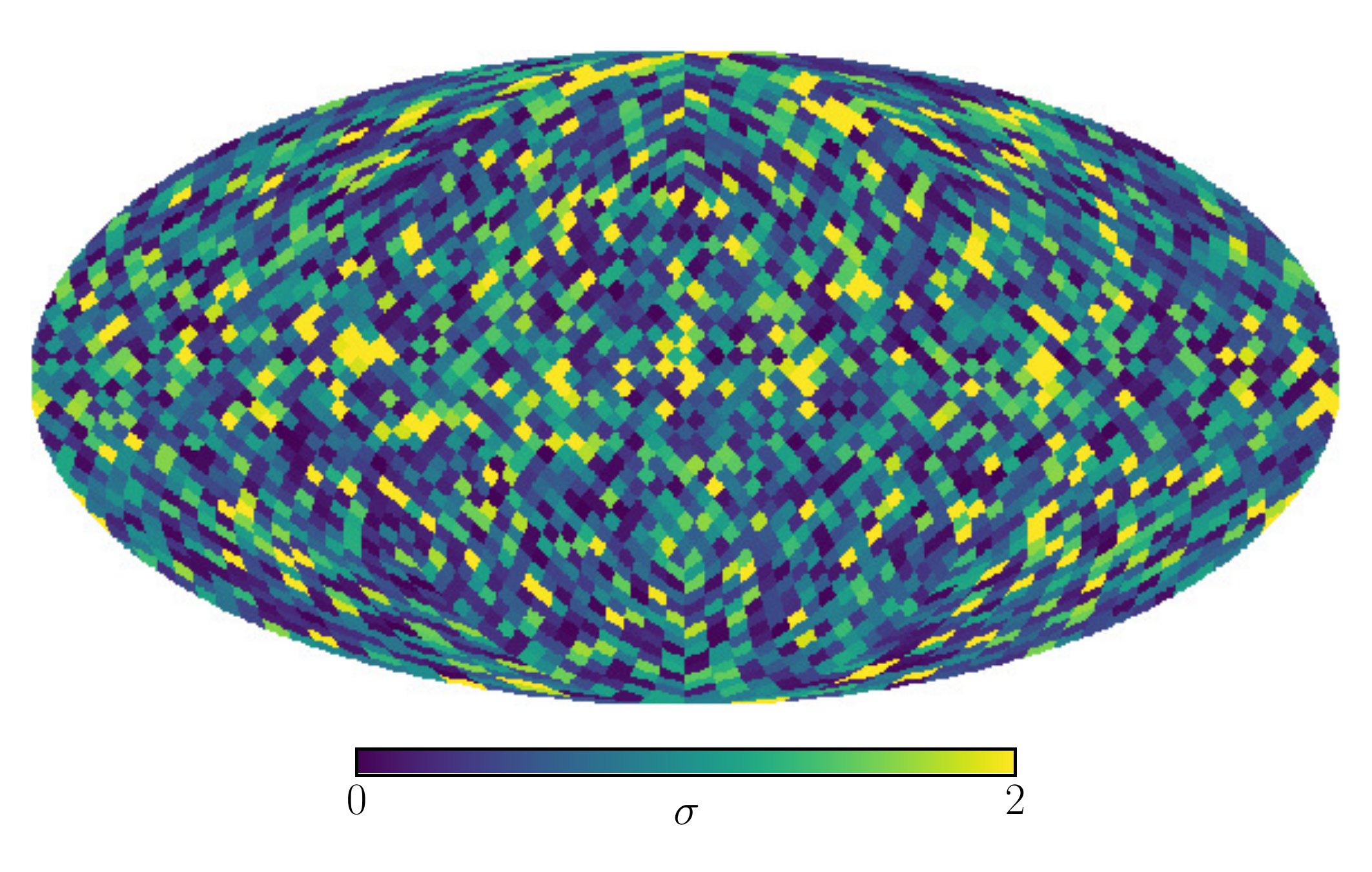}\\
    \includegraphics[width=0.4\textwidth]{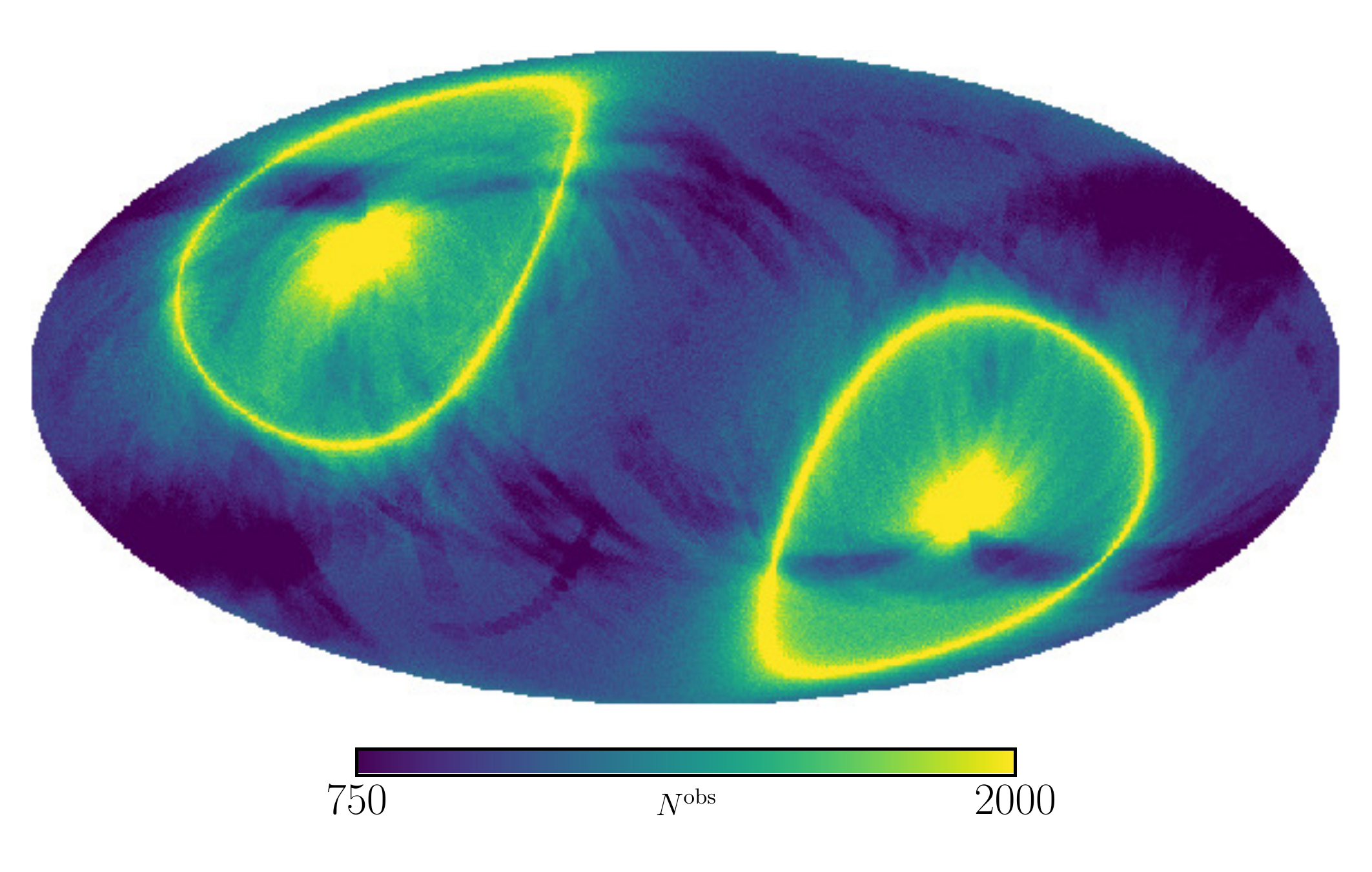}
    \caption{Top and middle panel:  Same as Fig.~\ref{fig:A3}, just for skewness and kurtosis weighted by $N^{\rm obs}$ as defined in Eqs.~\ref{eq:B1}--\ref{eq:B4}.  Bottom panel:  The number of observations $N^{\rm obs}$.}
    \label{fig:B1}
\end{figure}

In order to underpin the skewness/kurtosis results of section~\ref{sec:4.1} with regards to noise in the maps, we here repeat the calculations after down-weighting pixels with lower reliability.  We do this by example of the WMAP K-band $Q$ and $U$ maps which come with a maps of $N^{\rm obs}$, the number of observations per pixel.  We compute the weighted and standardized skewness and excess kurtosis in the standard way:

\begin{align}
&\gamma^{\rm w}_1 = \frac{V_1}{V_1^2-V_2}\sum_i^NN^{\rm obs}_i\left(\frac{x_i-m^{\rm w}}{\sigma^{\rm w}}\right)^3\label{eq:B1}\\
&\gamma^{\rm w}_2 = \frac{V_1}{V_1^2-V_2}\sum_i^NN^{\rm obs}_i\left(\frac{x_i-m^{\rm w}}{\sigma^{\rm w}}\right)^4-3,
\label{eq:B2}
\end{align}
where $V_j\equiv\sum_i^N(N^{\rm obs}_i)^j$, and the weighted mean and weighted standard deviation compute as
\begin{align}
&m^{\rm w} = \frac{1}{V_1}\sum_i^NN^{\rm obs}_ix_i\label{eq:B3}\\
&\left(\sigma^{\rm w}\right)^2 = \frac{V_1}{V_1^2-V_2}\sum_i^NN^{\rm obs}_i(x_i-m^{\rm w})^2.\label{eq:B4}
\end{align}
The significance maps which are shown in Fig.~\ref{fig:B1} should be compared with those of Fig.~\ref{fig:A3}:  hardly any difference can be seen by eye.  The reason is to be found in the largely smooth sky coverage of the satellite, i.e.~the map of $N^{\rm obs}$ changes little below the scale of one patch.  This also holds for Planck's 353~GHz map, even though, due to the lack of a $N^{\rm obs}$ product, one must estimate corresponding weights from the data.

%%%%%%%%%%%%%%%%%%%%%%%%%%%%%%%%%%%%%%%%%%%%%%%%%%

\bsp
\label{lastpage}
\end{document}